\pdfoutput=1

\RequirePackage{fix-cm}
\RequirePackage{rotating}

\documentclass[sigconf]{acmart}
\AtBeginDocument{%
  \providecommand\BibTeX{{%
    \normalfont B\kern-0.5em{\scshape i\kern-0.25em b}\kern-0.8em\TeX}}}

\usepackage[flushleft]{threeparttable}
\usepackage{multirow}
\usepackage{xcolor}
\usepackage{graphicx}
\usepackage{caption}
\usepackage{subcaption}
\usepackage[skins]{tcolorbox}
\usepackage{soul}
\usepackage{amsmath}
\usepackage[linesnumbered,ruled,vlined]{algorithm2e}
\usepackage{blindtext}
\usepackage[tikz]{bclogo}
\usepackage{color, colortbl}

\newenvironment{RQ}{\vspace{2mm}\begin{tcolorbox}[enhanced,width=3.33in,size=fbox,%fontupper=\small,
colback=blue!5,drop shadow southwest,sharp corners]}{\end{tcolorbox}}

\copyrightyear{2022}
\acmYear{2022}
\setcopyright{acmcopyright}\acmConference[MSR '22]{19th International Conference on Mining Software Repositories}{May 23--24, 2022}{Pittsburgh, PA, USA}
\acmBooktitle{19th International Conference on Mining Software Repositories (MSR '22), May 23--24, 2022, Pittsburgh, PA, USA}
\acmPrice{15.00}
\acmDOI{10.1145/3524842.3528437}
\acmISBN{978-1-4503-9303-4/22/05}

\acmConference[MSR 2022]{MSR '22: Proceedings of the 19th International Conference on Mining Software Repositories}{May 23–24, 2022}{Pittsburgh, PA, USA}
\begin{document}

%%
%% The "title" command has an optional parameter,
%% allowing the author to define a "short title" to be used in page headers.
\title{Dazzle: Using Optimized Generative Adversarial Networks to Address Security Data Class Imbalance Issue}

%%
%% The "author" command and its associated commands are used to define
%% the authors and their affiliations.
%% Of note is the shared affiliation of the first two authors, and the
%% "authornote" and "authornotemark" commands
%% used to denote shared contribution to the research.
\author{Rui Shu, Tianpei Xia, Laurie Williams, Tim Menzies}
\affiliation{%
  \institution{North Carolina State University}
  \city{Raleigh}
  \state{North Carolina}
  \country{USA}}
%   \email{{rshu, txia4, lawilli3}@ncsu.edu, timm@ieee.org}

%%
%% By default, the full list of authors will be used in the page
%% headers. Often, this list is too long, and will overlap
%% other information printed in the page headers. This command allows
%% the author to define a more concise list
%% of authors' names for this purpose.

%%
%% The abstract is a short summary of the work to be presented in the
%% article.
\begin{abstract}
{\bf Background:} Machine learning techniques have been widely used and demonstrate promising performance in many software security tasks such as software vulnerability prediction. However, the class ratio within software vulnerability datasets is often highly imbalanced (since the percentage of observed vulnerability is usually very low). {\bf Goal:} To help security practitioners address software security data class imbalanced issues and further help build better prediction models with resampled datasets. {\bf Method:} We introduce an approach called \textit{\textbf{Dazzle}} which is an optimized version of conditional Wasserstein Generative Adversarial Networks with gradient penalty (cWGAN-GP). Dazzle explores the architecture hyperparameters of cWGAN-GP with a novel optimizer called Bayesian Optimization. We use Dazzle to generate minority class samples to resample the original imbalanced training dataset. {\bf Results:} We evaluate Dazzle with three software security datasets, i.e., Moodle vulnerable files, Ambari bug reports, and JavaScript function code. We show that Dazzle is practical to use and demonstrates promising improvement over existing state-of-the-art oversampling techniques such as SMOTE (e.g., with an average of about 60\% improvement rate over SMOTE in recall among all datasets). {\bf Conclusion:} Based on this study, we would suggest the use of optimized GANs as an alternative method for security vulnerability data class imbalanced issues.

\end{abstract}

%%
%% The code below is generated by the tool at http://dl.acm.org/ccs.cfm.
%% Please copy and paste the code instead of the example below.
%%
% \begin{CCSXML}
% <ccs2012>
%  <concept>
%   <concept_id>10010520.10010553.10010562</concept_id>
%   <concept_desc>Computer systems organization~Embedded systems</concept_desc>
%   <concept_significance>500</concept_significance>
%  </concept>
%  <concept>
%   <concept_id>10010520.10010575.10010755</concept_id>
%   <concept_desc>Computer systems organization~Redundancy</concept_desc>
%   <concept_significance>300</concept_significance>
%  </concept>
%  <concept>
%   <concept_id>10010520.10010553.10010554</concept_id>
%   <concept_desc>Computer systems organization~Robotics</concept_desc>
%   <concept_significance>100</concept_significance>
%  </concept>
%  <concept>
%   <concept_id>10003033.10003083.10003095</concept_id>
%   <concept_desc>Networks~Network reliability</concept_desc>
%   <concept_significance>100</concept_significance>
%  </concept>
% </ccs2012>
% \end{CCSXML}

% \ccsdesc[500]{Computer systems organization~Embedded systems}
% \ccsdesc[300]{Computer systems organization~Redundancy}
% \ccsdesc{Computer systems organization~Robotics}
% \ccsdesc[100]{Networks~Network reliability}

%%
%% Keywords. The author(s) should pick words that accurately describe
%% the work being presented. Separate the keywords with commas.

%%
%% This command processes the author and affiliation and title
%% information and builds the first part of the formatted document.
\settopmatter{printfolios=true}
\maketitle

\section{Introduction}

\begin{figure}[!b]
\centering
\includegraphics[width=8cm]{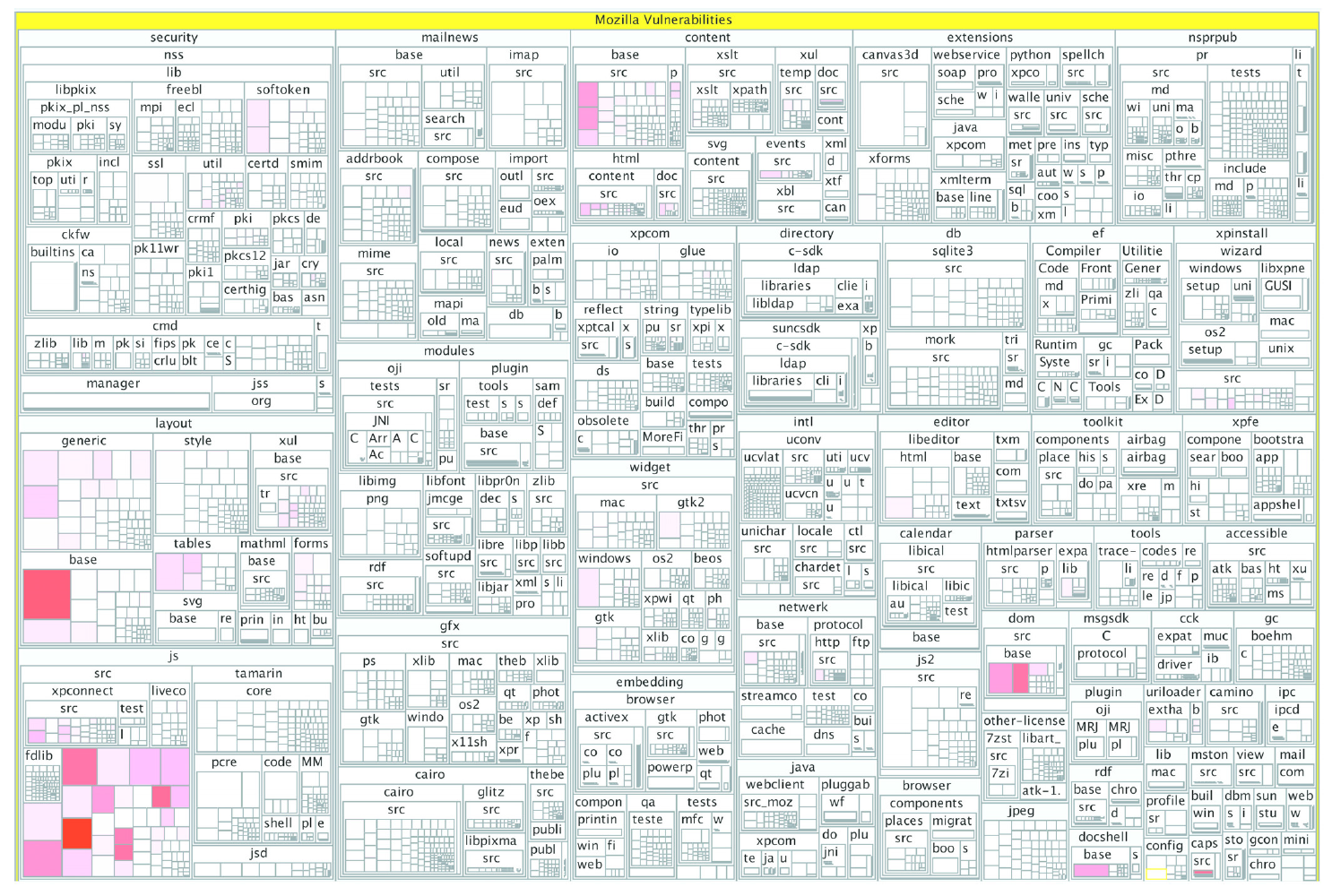}
\caption{ Mozilla code~\cite{neuhaus2007predicting}. Only a few modules  (seen in red) are vulnerable components. }
\label{fig:vulDistribution}
\end{figure}

Machine learning has been used for many security tasks; e.g. security vulnerability prediction~\cite{ghaffarian2017software}. A core problem with a security dataset is class imbalance; i.e., there may be very few instances of security events within many such datasets. For example, Figure~\ref{fig:vulDistribution} shows that components with known security vulnerabilities within Mozilla are very rare. As another example, as security bug reports can describe the critical security vulnerabilities in software products, Peters et al.~\cite{DBLP:journals/tse/PetersTYN19} show that only 0.8\% of bug reports are known to be security bug reports in their study.

\begin{table*}[!t]
\small
\centering
\caption{Recent works that use GANs as data oversampler in the security domain (as shown in column three, none of these prior works explored hyperparameter optimization on GANs). Training a GANs model is a difficult task since we have to achieve a balance between its internal components (i.e., the generator and discriminator)~\cite{saxena2021generative}. This paper explores this task with hyperparameter optimization with the novel Bayesian optimization.}
\begin{tabular}{c|l|c|l}
\hline
\textbf{Publication} & \textbf{Year} & \textbf{Optimized} & \multicolumn{1}{c}{\textbf{Brief Description}} \\ \hline \hline
\cite{anand2018phishing} & 2018 & No & Propose text-GANs to generate phishing URLs. \\ 
\cite{fiore2019using} & 2019 & No & Train a GANs to output mimicked minority class example for credit card fraud detection. \\ 
\cite{huang2020igan} & 2020 & No & Propose IGAN-IDS to cope with imbalanced intrusion detection. \\ 
\cite{wu2020using} & 2020 & No & Use an improved GANs to detect social bots on Twitter. \\ 
\cite{chen2021using} & 2021 & No & Use GANs as data augmentation in Android Malware Detection. \\ 
\cite{shi2018generative} & 2018 & No & Apply GANs for black-box API attacks to deal with limited training data. \\ 
\cite{cheng2019pac} & 2019 & No & Propose using CNN GANs to generate network traffic. \\ 
\cite{shahriar2020g} & 2020 & No & Propose a GANs based intrusion detection system to counter imbalanced learning. \\
\cite{yan2019automatically} & 2019 & No & Use GANs to synthesize DoS attack traces.  \\ 
\cite{ring2019flow} & 2019 & No & GANs is used to generate flow-based network traffic.\\ 
\cite{truong2020empirical} & 2020 & No & Adopt two existing GANs models to generate synthetic network traffic. \\
\end{tabular}
\label{tbl:dataAugumentation}
\end{table*}

When the target class is rare,   it is challenging for a   learner  to distinguish the goal (security target) from other  event~\cite{kaur2019systematic}. There are many ways to handle the class imbalance. For example, SMOTE (i.e. \ul{\textbf{S}}ynthetic \ul{\textbf{M}}inority \ul{\textbf{O}}versampling \ul{\textbf{TE}}chnique)~\cite{chawla2002smote} is a highly-cited methods that \textit{oversamples} the minority class by generating new samples. Specifically, SMOTE works by introducing new synthetic samples along with the line segments of $k$ nearest minority class neighbors. However, SMOTE generates new samples via a simplistic linear interpolation between minority neighbors. Also, when generating new data in some local regions, SMOTE does not use knowledge from the whole minority class samples -- which means its interpolations might not be helpful. Recently, SMOTE has been used extensively in software analytics in work published at top venues such as ICSE~\cite{agrawal2018better,tan2015online}, TSE~\cite{bennin2017mahakil}, EMSE~\cite{jiarpakdee2020impact}, etc.

SMOTE was first proposed in 2002, and this paper explores ``can we do better than SMOTE?''. For example, a new approach to generate samples for resampling purposes is GANs~\cite{DBLP:journals/corr/GoodfellowPMXWOCB14}; i.e. \ul{\textbf{G}}enerative \ul{\textbf{A}}dversarial \ul{\textbf{N}}etworks. Unlike SMOTE's issue with local inference, GANs oversampling can effectively learn the whole data characteristics and generate samples close to the distribution of original input data. Considering that GANs can achieve some impressive results in producing meaningful, realistic samples in prior studies (e.g., in domains such as computer vision~\cite{jabbar2021survey,wang2021generative}), more security practitioners have adopted variants of GANs in many security tasks~\cite{anand2018phishing,huang2020igan,cheng2019pac} (see Table~\ref{tbl:dataAugumentation}).

One reason to prefer SMOTE over GANs is that the SMOTE is much easier to implement and apply. GANs have two parts: a {\em generator} model that generates new plausible examples and a {\em discriminator} model that checks if it can distinguish real from fake examples. However, training a useful and \textit{stable} GANs can be a difficult task~\cite{saxena2021generative}. Here, {\em stable} means a balance between generator and discriminator with proper coordination. For example, if one model overpowers the other, neither can learn more even with more iterations. Some other challenges with training GANs include {\em mode collapse} (discussed in \S\ref{sec:challenge}), in which situation the generator may not explore much of the possible solution space and thus fails to produce a variety of realistic outputs.

This empirical study tries to tame the GANs training problem as well as using GANs as a data oversampler with {\em hyperparameter optimization} on Wasserstein GAN (WGAN)~\cite{DBLP:conf/iclr/ArjovskyB17,DBLP:conf/icml/ArjovskyCB17}. WGAN applies the Wasserstein distance metric instead of the cross-entropy loss used in the traditional discriminator. The advantage of the Wasserstein distance metric is that it measures the distributions of each data feature and determines how far apart the distributions are for real and fake data. Considering the complexity of tuning two components in the GANs architecture, we use a novel optimizer called \textit{Bayesian Optimization}~\cite{snoek2012practical,shahriari2015taking}. Our Bayesian optimizer explores the hyperparameter set of WGAN's generator and discriminator and returns an optimal solution set towards the evaluation target. We refer to our proposed combination of GANs and Bayesian Optimization as ``\textit{\textbf{Dazzle}}''. The experiments of this paper evaluate Dazzle with three security datasets, i.e., Moodle vulnerable files, Ambari bug reports, and JavaScript function code dataset. The results show that we can achieve an average 60\% improvement rate in recall across all datasets. We recommend using optimized GANs for security vulnerability dataset class rebalancing purposes based on this study.

As for the novelty and contribution of this work, we note that this paper is not the first work to apply Bayesian optimization to tune the GANs architecture. For example, prior work~\cite{elakkiya2021optimized} has proposed using optimized GANs in the sign language classification. However, the main focus of this empirical study is to show that the idea of using optimized GANs is able to help solve some existing security tasks, and it is more promising than currently widely used SMOTE-based methods. We also note that this study cannot cover all security tasks as we show in Table~\ref{tbl:dataAugumentation}, and we believe this would be an interesting future direction to explore.

The remainder of this paper is organized as follows. We discuss background and related work in Section~\ref{sec:background} and our methodology in Section~\ref{sec:methodology}. We then report our experiment details in Section~\ref{sec:Evaluation}, including datasets, evaluation metrics, etc. Section~\ref{sec:results} presents our experiment results. We discuss the threats of validity in Section~\ref{sec:threats} and provides a remark of addressing class imbalanced issues in Section~\ref{sec:discussion} and then we conclude in Section~\ref{sec:conclusion}.

\section{Background and Related Works}\label{sec:background}

\subsection{Software Vulnerability Prediction}

Software security vulnerabilities are critical issues that would impact software systems' confidentiality, integrity, and availability. The exploitation of such vulnerabilities would result in tremendous financial loss. To mitigate these issues, many machine learning and data mining techniques are proposed to build vulnerability prediction models to aid security practitioners~\cite{ghaffarian2017software}. 

Prior works have demonstrated several ways to extract useful features to train vulnerability prediction models. For example, as software bug reports can describe security vulnerabilities in software products, prior researchers~\cite{DBLP:journals/tse/PetersTYN19,jiang2020ltrwes,wu2021data} proposed a way to adopt natural language text based text mining techniques to identify security-related keywords. Vulnerability prediction models are then built by using the frequency of security-related keywords as features. Source code is another widely used avenue to derive vulnerability prediction models. Each piece of code can be represented by text, metric, token, tree, or graph. For example, in the metric-based representation, a code fragment is represented by a vector of features, such as lines of code, number of functions, total external calls, etc. These metrics can be extracted automatically with existing source code analyzers or extractors, which become ideal available resources to train prediction models~\cite{li2016vulpecker,walden2014predicting,ferenc2019challenging}. Metric-based representation is often used at the file/component fragment level.

\subsection{Software Vulnerability Dataset Class Imbalance}

Software bug report based~\cite{DBLP:journals/tse/PetersTYN19} or source code metric based~\cite{morrison2015challenges} prediction models mostly require a large amount of prior knowledge of vulnerabilities, which means many known vulnerable bug reports or codes are needed to train supervised machine learning models effectively. However, the imbalance between non-security bug reports and security bug reports or non-vulnerable code and vulnerable code brings significant challenges. When training machine learning prediction models with those class imbalanced datasets, the resulting models usually demonstrate a heavy bias towards the majority class. They tend to classify new data into the majority class, but they belong to the minority class. Such a phenomenon makes prediction models difficult to detect rare vulnerabilities (which are important) since models cannot effectively learn the decision boundary, resulting in poor performance.

Many prior studies have introduced various ways to tackle this issues, such as utilizing the ``sampling'' idea with the imbalanced data and they mainly fall into the following categories:

\begin{itemize}
    \item \textit{Undersampling} to remove majority class instances;
    \item \textit{Oversampling} to generate more of the minority class instances;
    \item Some \textit{hybrid} of the first two methods.
\end{itemize}

How to choose an appropriate way to sample the datasets is based on the characteristics of the datasets. Machine learning researchers~\cite{haixiang2017learning} advise that \textit{undersampling} usually works better than \textit{oversampling} if there are hundreds of minority instances in the datasets. When there are only a few dozen of minority instances, the \textit{oversampling} approaches are superior to \textit{undersampling}. In the case of large size training datasets, the \textit{hybrid} methods would be preferred. The datasets we studied fall into the second category. Therefore, \textit{oversampling} is a better choice.

\subsection{SMOTE}

A simple way to oversample data is to duplicate samples from the minority class in the training dataset before training a model. Samples from the training dataset are selected randomly with replacement. This method is called \textit{RandomOverSampler}. It is referred to as a ``naive'' method because it assumes nothing about the data and provides no additional information to the model but barely balances the class distribution.

% This means that examples from the minority class can be chosen and added to the new ``more balanced'' training dataset multiple times. In some cases, seeking a balanced distribution for a severely imbalanced dataset can cause affected algorithms to overfit the minority class, leading to increased generalization error. The effect can be better performance on the training dataset, but worse performance on the holdout or test dataset. The increase in the number of examples for the minority class, especially if the class skew was severe, can also result in a marked increase in the computational cost when fitting the model, especially considering the model is seeing the same examples in the training dataset again and again.

\begin{table*}[!htbp]
\small
\centering
\caption{A list of baseline data oversampling methods used in this study.}
\begin{tabular}{c|l}
\hline
\textbf{Method} & \multicolumn{1}{c}{\textbf{Description}} \\ \hline \hline
RandomOverSampler & Randomly duplicate examples in the minority class. \\ \hline
SMOTE & Create a synthetic sample between minority sample and its neighbour. \\ \hline
ADASYN & Creates synthetic data according to the data density. \\ \hline
BorderlineSMOTE & Only select minority samples that are misclassified. \\ \hline
KMeansSMOTE & Apply a KMeans clustering before to over-sample using SMOTE. \\ \hline
SVMSMOTE & Use an SVM algorithm to detect sample to use for generating new synthetic samples. \\ \hline
SMOTUNED & An auto tuning version of SMOTE that optimizes its parameters. \\ \hline
\end{tabular}
\label{tbl:smoteVairant}
\end{table*}

\begin{figure}[!tp]
\centering
\includegraphics[width=6.0cm]{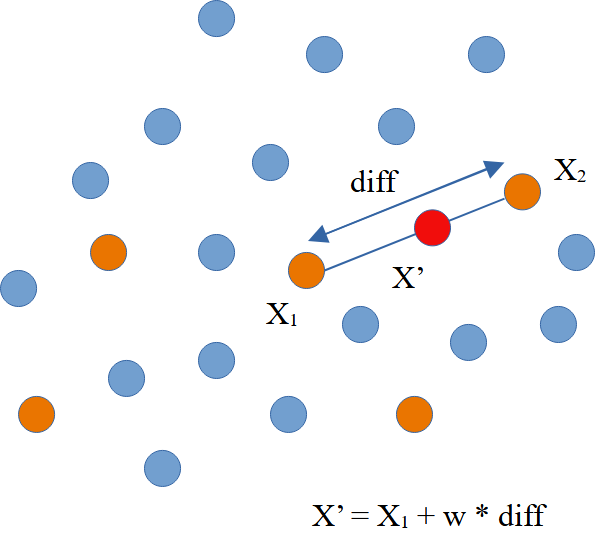}
\caption{An example of how SMOTE works. The blue dots denote the majority class samples and the orange dots denote the minority class samples. In SMOTE, a neighbour sample $X_{2}$ is selected for sample $X_{1}$ and a new synthetic sample $X'$ (i.e., the red dot) is created as a linear interpolation.}
\label{fig:smotearchitecture}
\end{figure}

An improved way is synthesizing new samples with existing samples from the minority class. \ul{\textbf{S}}ynthetic \ul{\textbf{M}}inority \ul{\textbf{O}}versampling \ul{\textbf{TE}}chnique, also known as \textbf{SMOTE}~\cite{chawla2002smote}, is an algorithm that oversamples the minority class by creating new synthetic samples. SMOTE works by selecting samples that are close in the feature space, drawing a line between the samples, and generating a new sample at a point along that line (as shown in Figure~\ref{fig:smotearchitecture}). Specifically, SMOTE calculates the $k$ nearest neighbors for each minority class sample. Depending on the amount of oversampled instances required, one or more of the $k$-nearest neighbors are selected to create the synthetic samples. This amount is usually denoted by oversampling percentage (e.g., $50\%$ by default). The next step is to create a synthetic sample connecting two minority samples randomly.

Algorithm~\ref{alg:smote} describes how SMOTE works. A random sample from the minority class is firstly chosen. Then $k$ of the nearest neighbors of that example are found. For each selected neighbor, a synthetic example is created at a randomly selected point between the two samples in feature space. The approach is more effective than the naive duplicate oversampling because new synthetic samples from the minority class are created that are plausible and relatively close in feature space to existing samples from the minority class.

%%% Coloring the comment as blue
\newcommand\mycommfont[1]{\footnotesize\ttfamily\textcolor{blue}{#1}}
\SetCommentSty{mycommfont}

\begin{algorithm}[t]
    \SetKwInOut{Input}{Input}
    \SetKwInOut{Output}{Output}

    \underline{\textbf{Function} SMOTE} $(D_{training}, k, m, r)$\;
    \Input{Training datasets - $D_{\mathit{training}}$, \\ 
            Number of nearest neighbours - $k$, \\ 
            Number of synthetic instances to create - $m$, \\
            Distance metric parameter - $r$}
    \Output{Resampled training datasets - $D_{\mathit{resampled}}$}
    \While{\# of Minority samples $<$ $m$}
    {
     $x$ $\leftarrow$ random minority class samples from $D_{\mathit{training}}$ \\
     $\mathit{neighbours}$ $\leftarrow$ $k$ nearest neighbours of $x$ \\
     
     \For{$n_{i}$ $\in$ $\mathit{neighbours}$}    
        { 
        	$x_{\mathit{new}}$ $\leftarrow$ $\mathit{interpolate}(x, n_{i})$ \\
        	Add $x_{\mathit{new}}$ to $D_{\mathit{resampled}}$
        }
    }
    \Return $D_{\mathit{resampled}}$
    \caption{Pseudocode for SMOTE.}
    \label{alg:smote}
\end{algorithm}

Table~\ref{tbl:smoteVairant} also lists several variants of SMOTE, which are used as our baseline methods for comparison purposes. For example, \textit{ADASYN}~\cite{he2008adasyn} (i.e., Adaptive Synthetic Sampling) is an improved version of SMOTE, which creates synthetic data according to the data density. The synthetic data generation would be inversely proportional to the density of the minority class. It means more synthetic data are created in regions of the feature space where the density of minority examples is low and fewer or none where the density is high. \textit{BorderlineSMOTE}~\cite{han2005borderline} involves selecting those instances of the minority class that are misclassified. Unlike with the SMOTE, where the synthetic data are created randomly between the two data, BorderlineSMOTE only makes synthetic data along the decision boundary between the two classes. \textit{KMeansSMOTE}~\cite{last2017oversampling} applies a KMeans clustering before to over-sample using SMOTE, and \textit{SVMSMOTE}~\cite{nguyen2011borderline} uses an SVM algorithm to detect samples to use for generating new synthetic samples. \textit{SMOTETUNED}~\cite{agrawal2018better} is an auto-tuning version of SMOTE that explores the parameter space of SMOTE with an optimizer called differential evolution algorithm.

% \begin{algorithm}[!t]
% \small
% \hspace{0.2cm}\begin{lstlisting}[xrightmargin=5.0ex,mathescape,frame=none]
% def SMOTE(k=2, m=50%, r=2): # defaults
%     while Minority < m do
%         add synthetic samples
        
% def synthetic_samples(X0):
%     relevant = emptySet
%     k1 = 0
%     while(k1++ < 20) and size(relevant) < k {
%         all = k1 nearest neighbors of X0
%         relevant += item in all of X0 class
%     }
%     Z = any of relevant
%     Y = interpolate(X0, Z)
%     return Y
    
% def minkowski_distance(a, b, r):
%     return ($\Sigma_{i}abs(a_{i}-b_{i})^{r})^{1/r}$

% \end{lstlisting}
% \caption{Pseudocode of SMOTE From~\cite{agrawal2017better}.}\label{algorithm:smote}  
% \end{algorithm}

% Please add the following required packages to your document preamble:
% \usepackage{multirow}

\begin{figure}[t]
\centering
\includegraphics[width=9cm]{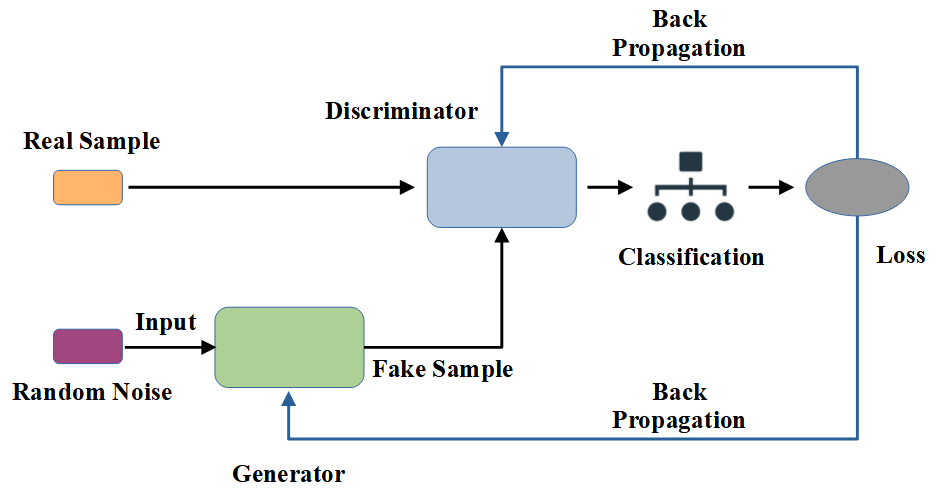}
\caption{The architecture of a traditional GANs model.}
\label{fig:ganarchitecture}
\end{figure}

\subsection{GANs}\label{gan101}

Compared with SMOTE, GANs is a new emerging technique, and in this work, we explore the merits of GANs over SMOTE. \ul{\textbf{G}}enerative \ul{\textbf{A}}dversarial \ul{\textbf{N}}etworks (\textbf{GANs})~\cite{DBLP:journals/corr/GoodfellowPMXWOCB14} are a neural network architecture that has a set of two models used to produce synthetic data. The GANs model architecture (see Figure~\ref{fig:ganarchitecture}) typically involves two sub-models, i.e., a \textit{generator} and a \textit{discriminator}. The generator model generates new plausible examples in the problem domain, while the discriminator model distinguishes whether the new generated examples by generator are real or fake, from the perspective of the domain. Both of the models are trained in a \textit{min-max zero-sum game} since the generator tries to produce synthetic instances of data that reliably trick the discriminator, while the discriminator tries to distinguish between real and fake data.

The two models, the generator and discriminator, are trained together. The generator generates a batch of samples, and these, along with real examples from the domain, are provided to the discriminator and classified as real or fake. The discriminator is then updated to get better at discriminating real and fake samples in the next round, and importantly, the generator is updated based on how well or not, the generated samples fooled the discriminator. When training begins, the generator produces obviously fake data, and the discriminator quickly learns to tell that it's fake. Finally, if generator training goes well, the discriminator gets worse at telling the difference between real and fake. It starts to classify fake data as real, and its accuracy decreases. Both the generator and the discriminator are neural networks. The generator output is connected directly to the discriminator input. Through backpropagation, the discriminator's classification provides a signal that the generator uses to update its weights. In fact, a really good generative model may be able to generate new examples that are not just plausible, but indistinguishable from real examples from the problem domain.

The loss function of GANs is shown as follows:

\begin{equation}
  \min\limits_{G}\max\limits_{D}V(D,G) = E_x[log(D(x))] + E_z[log(1-D(G(z))]
\end{equation}
where $D(x)$ is the discriminator's estimate of the probability that real data instance $x$ is real, $E_{x}$ is the expected value over all real data instance. $G(x)$ is the generator's output when given noise $z$ and $D(G(z)$ is the discriminator's estimate of the probability that a fake instance is real. $E_{z}$ is the expected value over all random inputs to the generator. The goal of discriminator is to bring $D(G(Z))$ closer to 0, while the goal of generator is to bring it closer to 1. If the generator outputs a probability of 0.5, then this means the discriminator is unable to make a right decision whether the instance is real or fake.

\begin{algorithm}[!t]
\footnotesize
    \SetKwInOut{Input}{Input}
    \SetKwInOut{Output}{Output}

    \underline{\textbf{Function} simpleGAN} $(D_{\mathit{training}}, D, G)$\;
    \Input{Training datasets - $D_{\mathit{training}}$, \\
            Discrinimator - $D$, \\
            Generator - $G$}
    \Output{Resampled training datasets - $D_{\mathit{resampled}}$}
    \For{$\mathit{epoch}_{i}$ $\in$ number of epochs}    
    { 
        \tcc{Train Discriminator $D$}
        Sample a mini-batch of real data, train as \textit{true} \\
        sample a mini-batch of fake data from Generator $G$, train as \textit{false} \\
        Update the gradient of Discriminator $D$ \\
        
        \tcc{Train Generator $G$}
        Sample a mini-batch of fake data from Generator $G$, which should be classified as \textit{true} \\
        Update the gradient of Generator $G$
    }
    
    Generate new data $X_{\mathit{new}}$ with Generator $G$ \\
    Add $X_{\mathit{new}}$ to $D_{\mathit{resampled}}$ \\
    \Return $D_{\mathit{resampled}}$
    \caption{Pseudocode for a simple GANs as an oversampler.}
    \label{alg:simpleGAN}
\end{algorithm}

GANs are rapidly evolving fields, delivering promising results in generating realistic examples across a range of problem domains, most notably in images tasks such as synthesizing images from text description~\cite{zhang2017stackgan}, image compression~\cite{agustsson2019generative}, image classification~\cite{zhu2018generative}, etc. In the security domain, prior work indicate that GANs would be an ideal technique to train a classification model to explore unforeseen data threats with generated data. Table~\ref{tbl:dataAugumentation} lists recent work that use GANs as data oversampler (used in a way similar to Algorithm~\ref{alg:simpleGAN}). Those works motivate our study, however, we also note that they hardly introduce any way to optimize their GANs architecture as we do in this study.

% (GANs)~\cite{DBLP:journals/corr/GoodfellowPMXWOCB14} are a neural network architecture that comprise a generator and a discriminator. The idea behind GANs can be summarized as training the generator and discriminator simultaneously, while the generator tries to confuse the discriminator by generating realistic data and the discriminator learns to classify the real and generated data as genuinely as possible. Eventually, an ideal GANs achieves a balance between two networks, as generator produces realistic data, and discriminator can better distinguish realistic and fake data.

\subsection{Challenges with traditional GANs}\label{sec:challenge}

Although GANs has achieved notable success in multiple domains, GANs also face several challenges which may cause issues such as \textit{unstable training}~\cite{jabbar2021survey,saxena2021generative}. 

\textit{Nash Equilibrium.} Nash Equilibrium (NE)~\cite{nash1951non} is a notion in game theory where two players come to a joint strategy in which each player select a best response (i.e., a strategy that yields the best payoff against the strategies chosen by the other player). In the context of GANs, the generator and discriminator represent the two players, which work in an adversarial way against each other. The generator and discriminator train themselves simultaneously for NE. When both generator and discriminator update their cost function independently without coordination, it is hard to achieve NE.

\textit{Vanishing Gradient.} Vanishing gradient occurs when one part of GANs is more powerful than the other part. For example, if the generator model is very poor, then the discriminator can easily distinguish between real and fake samples. This further causes the probability of the generated samples being real from generator close to zero, i.e., gradients of $log(1-D(G_{z}))$ will be very small. Therefore, discriminator fails to provide gradients and the generator will stop updating. 

\textit{Mode Collapse.} Model collapse is one of the most crucial issues with GANs training, which means the output samples from generator lacks of variety (i.e., producing same outputs). If the generator starts to produce the same output, an ideal strategy for the discriminator is to reject the output. However, if the discriminator gets stuck in local minima and does not find the strategy, then the generator tends to find the same output that seems most plausible to the discriminator.

\subsection{Attempts to Address the Challenges}

Prior work indicates that Wasserstein GAN (WGAN)~\cite{DBLP:conf/iclr/ArjovskyB17,DBLP:conf/icml/ArjovskyCB17} is designed to prevent vanishing gradients. In WGAN, the discriminator does not classify input instances, but it outputs an exact score for each instance. WGAN does not use a threshold to decide whether an instance is real or fake but tries to make the score bigger than fake instances. WGAN also alleviates mode collapse since it prevents the discriminator from getting stuck in local minima. In this case, the generator has to try new samples since the discriminator would reject the same sample. For the Nash equilibrium problem (i.e., non-converge), prior work~\cite{farnia2020gans} suggests an exhaustive hyperparameter and architecture search, and hence this work. We will discuss WGAN and architecture optimization in detail in the next subsections.

\subsection{cWGAN-GP}\label{sec:cWGAN-GP}

Traditional GANs is motivated to minimize the distance between the actual and predicted probability distributions for real and generated samples. Typically, there are two metrics to measure the similarity between two probability distributions, the \textit{Kullback-Leibler divergence} and the the \textit{Jensen-Shannon divergence}.

\textit{Kullback-Leibler divergence}~\cite{kullback1997information}, also known as \textit{KL divergence}, is a metric to measure \textit{relative entropy} between two probability distributions over the same variable. Consider distributions $P$ and $Q$ of a continuous random variable, the KL divergence is computed as an integral as follows:

\begin{equation}
    KL(P\|Q) = \int p(x)log(\frac{p(x)}{q(x)}) \,dx
\end{equation}
where $p(x)$ and $q(x)$ are the probability density functions of distribution $P$ and $Q$, respectively. The lower the KL divergence value, the closer the two distributions are to each other.

% the KL divergence is not symmetrical. a divergence is a scoring of how one distribution differs from another, where calculating the divergence for distributions P and Q would give a different score from Q and P.

An extension to KL divergence is the \textit{Jensen-Shannon divergence}~\cite{fuglede2004jensen}, also known as \textit{JS divergence}. Compared with KL divergence, this metric is a symmetric version, which means calculating the divergence for distribution $P$ and $Q$ will result in the same score as from distribution $Q$ and $P$. Define the quantity $M = (P + Q)*0.5$, JS divergence is formulated as follows: 

\begin{equation}
    JS(P\|Q) = \frac{1}{2}KL(P\|M) + \frac{1}{2}KL(Q\|M)
\end{equation}
Besides symmetric, JS divergence is also a smoothed and normalized version, and the square root of this score which referred as \textit{Jensen-Shannon distance} is commonly used.

% It is more useful as a measure as it provides a smoothed and normalized version of KL divergence, with scores between 0 (identical) and 1 (maximally different), when using the base-2 logarithm.
% The square root of the score gives a quantity referred to as the Jensen-Shannon distance, or JS distance for short.

The JS divergence scores provides ways to calculate scores for cross-entropy which is commonly used as a loss function in classification models such as the discriminator in GANs. However, researchers notice that such loss function does not necessarily correlate with the sample quality and therefore does not guarantee the convergence between generator and discriminator to an equilibrium~\cite{goodfellow2016nips}. Wasserstein GAN~\cite{DBLP:conf/iclr/ArjovskyB17,DBLP:conf/icml/ArjovskyCB17} improves traditional GANs' optimization goal based on Wasserstein distance, which is formulated as follows:

\begin{equation}
    W(P, Q) = \inf\limits_{\gamma\sim\prod(P, Q)}E_{(x,y)\sim\gamma}[\|x - y\|]
\end{equation}
where $\prod(P, Q)$ is the set of all possible joint distributions in which $P$ and $Q$ are combined. For each possible joint distribution $\gamma$, a real sample $x$ and a generated sample $y$ can be sampled, and the sample distance $\|x - y\|$ is calculated, so that the expected value $E_{(x,y)\sim\gamma}[\|x - y\|]$ of the sample to the distance under the joint distribution $\gamma$ can be calculated. This expected value can be taken to the lower bound in all possible joint distributions and defined as the Wasserstein distance of the two distributions. This distance is helpful when facing two distributions with non-overlapping, in which case JS divergence fails to provide a useful gradient.

% Wasserstein distance (also known as Earth Mover's distance, a.k.a, EM distance) is a measure of the distance between two probability distributions. 
% Unlike KL divergence and JS divergence, the Wasserstein distance is a true probability metric and considers both the probability of and the distance between various outcome events. Wasserstein distance also provides a meaningful and smooth representation of the distance between distributions. 

% When facing two distributions with non-overlapping, JS divergence indicate infinite while the Wasserstein distance can indicate which two of the samples is closer. This helps to provide powerful gradients to GANs generator even when the quality of the generated samples is still poor.

WGAN with Gradient Penalty (WGAN-GP) further suggests to add a gradient penalty to address the concern of Lipschitz constraint. With the gradient penalty, the norm of the gradient is limited to a value of 1 to satisfy the 1-Lipschitz continuous condition. This is helpful to build a ``worse'' discriminator, but provide more gradient information that helps to train a better generator. In short, the use of gradient penalty helps enhance the training stability and reduce the mode collapse of the networks.

Moreover, we adopt an extension to WGAN-GP, which is called conditional WGAN-GP. In this method, both the generator and discriminator add data category information, with which the optimization function of WGAN-GP is a maximal and minimal game with this condition.

\subsection{Bayesian Optimization}~\label{sec:bayesian}

Since training a new GANs model can be difficulty, this works checks if that process can be automated with hyperparameter optimization. Typically a hyperparameter has a known effect on a model in the general sense, but it is not clear how to best set a hyperparameter for a given dataset. Hyperparameter optimization or hyperparameter tuning is a technique that explores a range of hyperparameters and search for the optimal solution for a task. \textit{Bayesian optimization}~\cite{snoek2012practical,shahriari2015taking} is a widely-used hyperparameter optimization technique that keeps track of past evaluation results. The principle of Bayesian optimization is using those results to build a probability model of objective function, and map hyperparameters to a probability of a score on the objective function, and therefore use it to select the most promising hyperparameters to evaluate in the true objective function. This method is also called \textit{Sequential Model-Based Optimization} (SMBO).

\begin{figure}[!tp]
 \includegraphics[width=8cm]{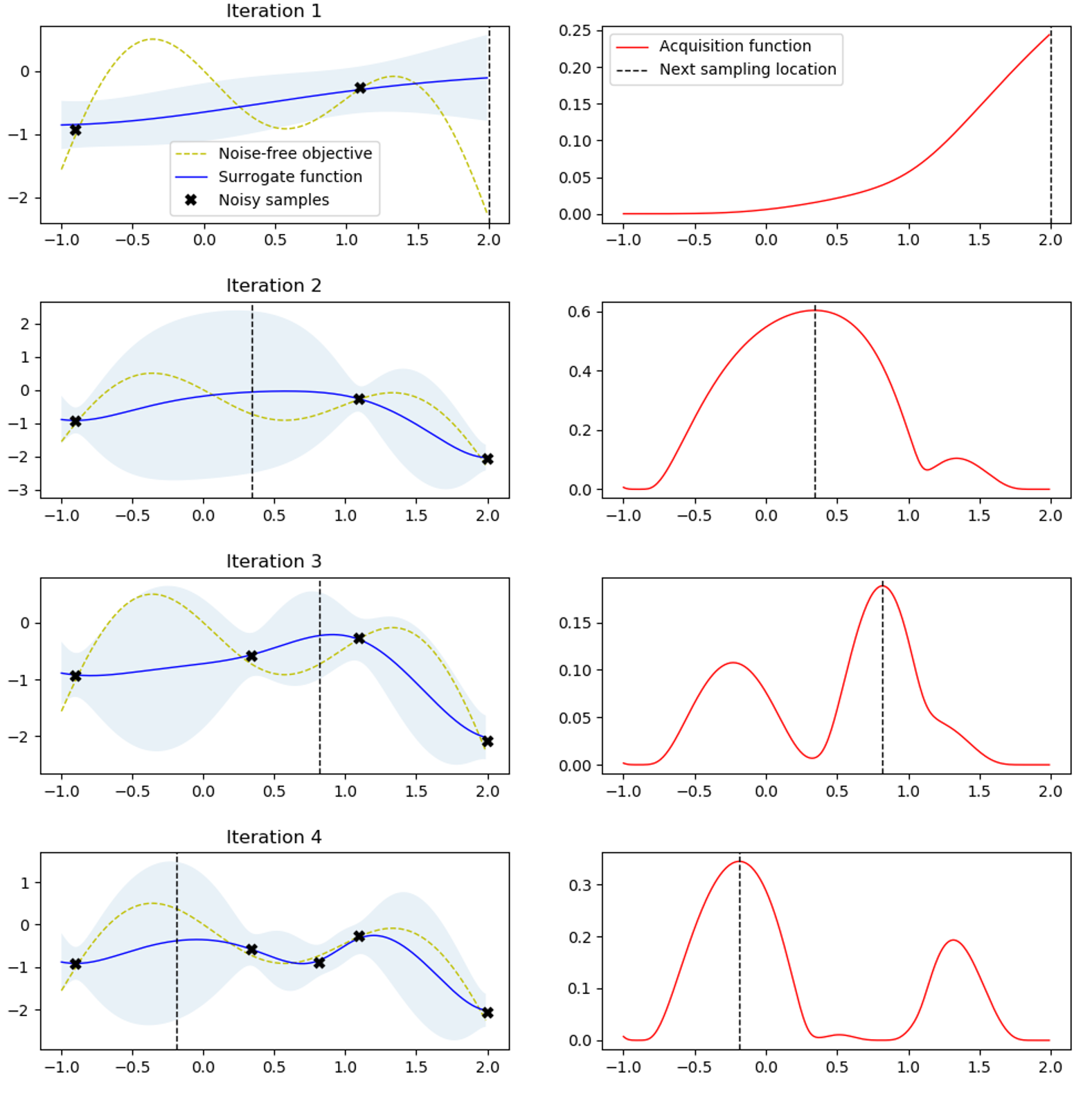}
\caption{An example of the Bayesian optimization process. Bayesian optimization incorporates prior belief about \textit{objective function} and updates the prior with samples drawn from objective function to get a posterior that better approximates objective function. The model used for approximating the objective function is called \textit{surrogate function}. Bayesian optimization also uses an \textit{acquisition function} that directs sampling to areas where an improvement over the current best observation is more likely. Note that the Bayesian optimization we use come from the HyperOpt~\cite{bergstra2015hyperopt} library, in which the optimization algorightm is based on Tree of Parzen Estimators (TPE).}
\label{fig:bayesian}
\end{figure}

The probability representation of the objective function is called \textit{surrogate function} or \textit{response surface} because it is a high-dimensional mapping of hyperparameters to the probability of a score on the objective function. The surrogate function is much easier to optimize than the objective function and Bayesian methods work by finding the next set of hyperparameters to evaluate the actual objective function by selecting hyperparameters that perform best on the surrogate function. This method continually updates the surrogate probability model after each evaluation of the objective function.

Several prior works have combined Bayesian optimization with GANs in tasks from other domains~\cite{dabayesian}~\cite{elakkiya2021optimized}, and this work shares the similar underlying idea with previous studies and adopts this combination in selected security tasks.

\section{Methodology}\label{sec:methodology}

\subsection{Dazzle: Optimized cWGAN-GP} 

In designing the network architecture of cWGAN-GP, another concern emerges as how to select the hyperparameters of the structure. GANs models might be highly sensitive to the hyperparameter selection. Prior work on DCGANs~\cite{DBLP:journals/corr/RadfordMC15} introduced a deep convolutional generative adversarial networks that made several modifications to the model hyperparameters of CNN architecture to address the architectural topology constraints and made the GANs' training more stable. For example, that work

\begin{enumerate}
    \item replaced pooling layers with strided convolutions and fractional-strided convolutions;
    \item used batch normalization for  generator \& discriminator;
    \item used ReLU activation in generator;
    \item used LeakyReLU activation in discriminator.
\end{enumerate}

\begin{table*}[!htbp]
\small
\centering
\caption{Hyperparameter selection ranges chosen to optimize in Dazzle.}
\begin{threeparttable}
\begin{tabular}{p{2in}|p{4in}}
\hline
\textbf{Hyperparameter} & \textbf{Range}\\ \hline \hline
Batch Size & 16, 32, 64, 128 \\ 
Learning Rate for Generator & 0.0005, 0.001, 0.005, 0.01, 0.05, 0.1  \\ 
Learning Rate for Discriminator & 0.0005, 0.001, 0.005, 0.01, 0.05, 0.1  \\ 
Optimizer for Generator & Adadelta, Adagrad, Adam, Adamax, NAdam, RMSprop, SGD \\ 
Optimizer for Discriminator & Adadelta, Adagrad, Adam, Adamax, NAdam, RMSprop, SGD \\ 
Activation Function for Generator & elu, relu, selu, sigmoid, softmax, tanh, hard\_sigmoid, softplus, leakyRelu \\ 
Activation Function for Discriminator & elu, relu, selu, sigmoid, softmax, tanh, hard\_sigmoid, softplus, leakyRelu \\
No. of Epochs & quniform(5, 20, 1)\\ 
Generator Layer Normalization & True, False  \\  
Discriminator Layer Normalization & True, False \\  
\end{tabular}
\begin{tablenotes}
    %   \small
\item * Note: \textbf{quniform($low$, $high$, $q$)} is a function returns a value like $round(uniform(low, high)/q) * q$, while \textbf{uniform($low$, $high$)} returns a value uniformly between $low$ and $high$. We also note that we do not tune the number of layers but with a fixed number (e.g., 4) in our study, and we find that such architecture suffices to achieve good performance on considered datasets.
% \item * Note-2: We do not optimize the number of hidden layers but use 3 layers in our study, since we observe that they are enough for us to make good predictions and speed up the training process.
\end{tablenotes}
\end{threeparttable}
\label{tbl:DazzleParameter}
\end{table*}

Inspired by this work, we hypothesize that \textit{GANs would benefit from an automatic optimized architecture}. We mean proper hyperparameter selection would help with GANs training to converge and further achieve better performance. In our case as using GANs as data oversampler, if we indicate a not well-designed GANs as GANs $A$ and a well-designed GANs as GANs $B$, then if we build classification model with training data from GANs $B$, then the prediction performance is better than the models built from data with GANs $A$.

\begin{algorithm}[!tb]\footnotesize
    \SetKwInOut{Input}{Input}
    \SetKwInOut{Output}{Output}

    \underline{\textbf{Function} Dazzle} $(D_{\mathit{training}}, D_{\mathit{validation}}, D, G, \theta, F )$\;
    \Input{Training datasets - $D_{\mathit{training}}$, \\
            Validation datasets - $D_{\mathit{validation}}$, \\
            Discrinimator - $D$, \\
            Generator - $G$, \\
            Hyperparameter space - $\theta$, \\
            Target function - $F$}
    \Output{Optimal resampled training dataset $D_{\mathit{resampled}_{\mathit{optimal}}}$, \\
    Optimal hyperparameter set $\theta_{\mathit{optimal}}$}
    \For{$\mathit{iteration}_{i}$ $\in$ number of Bayesian Optimization iterations}    
    { 
        Select a hyperparameter set $\theta_{i}$ $\in$ $\theta$ \\ 
        Train $D$ and $G$ with $\theta_{i}$ \\
        Generate new resampled training dataset $D_{\mathit{resampled}_i}$ \\
        Build classifier with $D_{\mathit{resampled_i}}$ and evaluate with $D_{\mathit{validation}}$ \\
        Compute loss with target function $F$ \\
    }
    Rank all optimization iterations by loss with smallest on the top \\
    \Return $D_{\mathit{resampled}_{\mathit{optimal}}}$ and $\theta_{\mathit{optimal}}$
    \caption{Pseudocode of Dazzle's training process.}
    \label{alg:dazzle}
\end{algorithm}

However, hyperparameter optimization is not a trivial work, especially when facing a complex system such as neural networks. The tuning process is more challenging since there are more hyperparameter with neural networks, and what's the most important, even one iteration of evaluation would be time consuming. Traditional hyperparameter optimization techniques such as ``random search'' or ``grid search'' either suffer from not ideal performance or would be costly expensive. To address this concern, we propose a method called \textit{\textbf{Dazzle}} that adopts a novel optimizer called Bayesian optimization that fine-tunes both generator model and discriminator model.  

Dazzle's training process is on the training dataset and validation dataset. During each iteration of Bayesian optimization, Dazzle selects a hyperparameters for discriminator and generator from Table~\ref{tbl:DazzleParameter}, and generates new minority samples. These samples are used to resample the original dataset (and     to build the classifier). Each time, the classifier is only evaluated with validation dataset. With the optimization goal, the loss is computed. Finally, we rank all the optimization iterations by loss with smallest on the top of the rank, and select the trained classifer from that iteration as the optimized classifier. Moreover, we choose 30 iterations for Bayesian optimization and repeat the whole experiment process 10 times.

Algorithm~\ref{alg:dazzle} lists the optimization steps of Dazzle. Note that our task with the security datasets is a binary classification problem. In Dazzle, we choose g-measure as our optimization goal (i.e., the target to increase). G-measure is the harmonic mean of recall and the complement of false positive rate. We choose g-measure based on the following considerations. For an imbalanced dataset where there is a skew in class distribution, we have two competing goals:
\begin{itemize}
    \item We  focus on minimizing false negatives, i.e., increase recall; 
    \item We prefer not to predict too many non-security samples as security samples, i.e., reduce false positive rate.
\end{itemize}
Therefore, g-measure is  ideal for chasing     both    goals.

% Learning rate
% Batch size
% Number of epochs
% Generator optimizer
% Discriminator optimizer
% Number of layers
% Number of units in a dense layer
% Activation function
% Loss function

\begin{table}[!t]
\centering
\small
\caption{Statistics of security datasets used in this study. Note that the security target column indicate the number of vulnerable files, security bug reports, and JavaScript function code, respectively.}
\begin{tabular}{c|c|c|c|c}
\hline
\textbf{Dataset} &
  \textbf{\begin{tabular}[c]{@{}c@{}}Security\\ Target\end{tabular}} &
  \textbf{Total} &
  \textbf{\begin{tabular}[c]{@{}c@{}}Imbalance\\ Rate (\%)\end{tabular}} &
  \textbf{\begin{tabular}[c]{@{}c@{}}No. of\\ Features\end{tabular}} \\ \hline
\begin{tabular}[c]{@{}c@{}}Moodle\\ Vulnerable Files\end{tabular}    & 24 & 2,942 & 0.8 & 13 \\ \hline
Ambari Bug Reports                                                          & 29 & 1,000 & 2.9 & 101 \\ \hline
\begin{tabular}[c]{@{}c@{}}JavaScript\\ Function Code\end{tabular} & 1,496 & 12,125 & 12.3 & 36 \\ \hline
\end{tabular}
\label{tbl:dataset}
\end{table}

\section{Experimental Evaluation}\label{sec:Evaluation}

\subsection{Datasets}

Our evaluations are experimented on datasets that are widely studied in prior work. \textbf{Moodle}~\cite{walden2014predicting} is an open source learning management system, and the data source for Moodle vulnerabilities is the National Vulnerabilities Database (NVD), from with a variety of vulnerabilities are covered, such as code injection, path disclosure, XSS, etc. A total of 24 vulnerable files are included in this dataset. \textbf{Ambari}~\cite{DBLP:journals/tse/PetersTYN19} is an open source project of Apache that aims to provision, manage and monitor Apache Hadoop cluster. Bug reports with BUG or IMPROVEMENT label from the JIRA bug tracking system are selected, and then the selected bug reports are further classified  with scripts or manually into six high impact bugs (i.e., Surprise, Dormant, Blocking, Security, Performance, and Breakage bugs). All the target bug reports in the Ambari dataset all belong to Security bug reports (i.e., bug reports of the type Security). The \textbf{JavaScript}~\cite{ferenc2019challenging} function code dataset extracts data from Node Security Platform and the Snyk Vulnerability Database, and used static source code metrics as predictor features. Table~\ref{tbl:dataset} shows a list and description of the datasets used in this study. As we can observe from the table, all datasets suffer from different levels of class imbalanced issues.

% \textbf{\textit{Twitter Spam Accounts.}}~\cite{chen20156}

% \textbf{\textit{Malicious URLs.}}~\cite{mamun2016detecting}. ISCX-URL2016

% \textbf{\textit{Vulnerable Functions.}} ~\cite{DBLP:conf/nips/ZhouLSD019} source code level functions with CGP representation.

% % Please add the following required packages to your document preamble:
% % \usepackage{multirow}
% \begin{table*}[!htbp]
% \centering
% \caption{Description of datasets.}
% \begin{tabular}{c|c|r|r|r|r|r}
% \hline
% \multirow{2}{*}{\textbf{Dataset}} & \multicolumn{1}{c|}{\multirow{2}{*}{\textbf{Features}}} & \multicolumn{2}{c|}{\textbf{Training}} & \multicolumn{2}{c|}{\textbf{Testing}} & \multicolumn{1}{c}{\textbf{Total}} \\ \cline{3-7} 
%  & \multicolumn{1}{c|}{} & \textbf{Benign} & \textbf{Malicious(\%)} & \textbf{Benign} & \textbf{Malicious(\%)} & \textbf{Both(\%)} \\ \hline \hline
%  Bug reports & Security Keywords & 478 & 22 & 493 & 7 & 1,000 (2.9\%) \\ 
%  JavaScript Vulnerability & Code Metrics & 8,503 & 1,197 & 2,126 & 299 & 12,125 (12.3\%) \\ 
%  &  &  &  &  &  &  \\ 
%  &  &  &  &  &  &  \\ 
%  &  &  &  &  &  &  \\ \hline
% \end{tabular}
% \label{tbl:dataset}
% \end{table*}

% \begin{table}[!htbp]
% \centering
% \caption{Different levels of data imbalacing defined in this study.}
% \begin{tabular}{c|l}
% \hline
% \textbf{Level} & \multicolumn{1}{c}{\textbf{Description}} \\ \hline \hline
% Slight &  \\ \hline
% Moderate &  \\ \hline
% Severe &  \\ \hline
% \end{tabular}
% \label{tbl:imbalanceLevel}
% \end{table}

\subsection{Machine Learning Algorithms}

We apply five machine learning algorithms, namely K-Nearest Neighbours (KNN), Logistic Regression (LR), Decision Tree (DT), Random Forest (RF) and Support Vector Machine (SVM) in our experiment. We choose them since they are widely used in previous literatures in different classification tasks in security~\cite{ucci2019survey} or other domains such as defect prediction~\cite{lessmann2008benchmarking}. We implement these algorithms with open source tool called Scikit-learn. In order to reduce the influence of model hyperparameters to our evaluation results, we adopt default settings from Scikit-learn. We do not claim that the list of algorithms that we use is complete, but we note that these algorithms are enough for our study purpose.

% \begin{itemize}
%     \item K Nearest Neighbors (KNN).
%     \item Logistic Regression (LR).
%     \item Decision Tree (DT).
%     \item Random Forest (RF).
%     \item Support Vector Machine (SVM).
%     \item LightGBM.
%     \item Adaboost.
%     \item GBDT.
% \end{itemize}

\subsection{Evaluation Metrics}

For the performance of the classification models, the confusion matrix is used, where TP, TN, FP and FN indicate true positive, true negative, false positive and false negative, respectively. We report the results of recall (\textit{pd}), false positive rate (\textit{pf}), f-measure and g-measure as we defined in Table~\ref{tbl:metric}. Note that precision and accuracy in the table are not endorsed in our study, since both of these metrics can be inaccurate for datasets where the positive class is rare case. For example, Menzies et al.~\cite{menzies2007problems} argue that when the target class is less than 10\%, the precision results become more a function of the random number generator used
to divide data (for testing purposes). G-measure is a composite metric, which is the harmonic mean of recall and the complement of false positive rate. A higher g-measure indicates higher recall and lower false positive rate. As we discuss before, this metric is also our optimization target in Dazzle. We also report f-measure for completeness purpose.

\begin{table}[!htbp]
\small
\centering
\caption{Performance evaluation metrics. Definitions of recall (pd), false positive rate (pf), precision (prec), accuracy (acc), f-measure (f1) and g-measure (g-score).}
\begin{tabular}{c|c}
\hline
\textbf{Metric} & \multicolumn{1}{c}{\textbf{Expression}} \\ \hline \hline
Recall (pd) & $\frac{TP}{TP + FN}$ \\ \hline
\begin{tabular}[c]{@{}c@{}}False Positive Rate\\ (pf)\end{tabular} &  $\frac{FP}{FP + TN}$\\ \hline
Precision (prec) & $\frac{TP}{TP + FP}$ \\ \hline
Accuracy (acc) & $\frac{TP + TN}{TP + TN + FP + FN}$ \\ \hline
F-Measure (f1) & $\frac{2 * pd * prec}{pd + prec}$ \\ \hline
G-Measure (g-score) & $\frac{2 * pd * (100 - pf)}{ pd + (100 - pf)}$ \\ \hline
% D2H & $\frac{\sqrt{(1 - pd)^{2} + (0 - pf)^{2}}}{\sqrt{2}}$ \\ \hline
\end{tabular}
\label{tbl:metric}
\end{table}

\subsection{Experiment Rigs}

Our datasets are split in a stratified way into two parts with a ratio of 8:2 where the latter part is used as testing set. We further split the former part into training set and validation set with the same ratio. Therefore, the final ratio between the actual training, validation, and testing part is 6.4: 1.6: 2 of the whole dataset. The training part is only used for training classifiers with selected hyperparameter set, and the validation part is used to evaluate the classifiers during optimization iterations towards optimization goal. 
Then the selected optimized models are evaluated on the testing dataset. 

Lastly, our implementation of Bayesian Optimization is based on the tool called Hyperopt~\cite{bergstra2013hyperopt}, which is one of the most cited hyperparameter optimizer in the literature at this time of writing. The implementation of SMOTE and its variants are based on the open-source imbalanced-learn toolbox~\cite{lemaitre2017imbalanced} while SMOTUNED is implemented according to Agrawal et al.'s study~\cite{agrawal2018better}. SMOTUNED has three available parameters:

\begin{itemize}
    \item Number of neighbours $k$ with range $[1, 20]$.
    \item Minkowski distance metric $r$ with range $[1, 6]$.
    \item Number of synthetic samples $m$ to create with range $[50, 500]$.
\end{itemize}

\section{Results}\label{sec:results}

\begin{table*}[!htbp]
\small
\centering
\caption{Median performance results (converted to range 0 - 100) from 10 repeats. Best performances are highlighted.}
\begin{tabular}{c|c|rrrrr||rrrrr||rrrrr}
\hline
\multirow{2}{*}{\textbf{Metric}} &
  \multirow{2}{*}{\textbf{Treatment}} &
  \multicolumn{5}{c|}{\textbf{\begin{tabular}[c]{@{}c@{}}Moodle\\ Vulnerable Files\end{tabular}}} &
  \multicolumn{5}{c|}{\textbf{\begin{tabular}[c]{@{}c@{}}Ambari\\ Bug Report\end{tabular}}} &
  \multicolumn{5}{c}{\textbf{\begin{tabular}[c]{@{}c@{}}JavaScript\\ Function Code\end{tabular}}} \\ \cline{3-17} 
 &
   &
  \multicolumn{1}{c|}{\textbf{KNN}} &
  \multicolumn{1}{c|}{\textbf{LR}} &
  \multicolumn{1}{c|}{\textbf{DT}} &
  \multicolumn{1}{c|}{\textbf{RF}} &
  \multicolumn{1}{c|}{\textbf{SVM}} &
  \multicolumn{1}{c|}{\textbf{KNN}} &
  \multicolumn{1}{c|}{\textbf{LR}} &
  \multicolumn{1}{c|}{\textbf{DT}} &
  \multicolumn{1}{c|}{\textbf{RF}} &
  \multicolumn{1}{c|}{\textbf{SVM}} &
  \multicolumn{1}{c|}{\textbf{KNN}} &
  \multicolumn{1}{c|}{\textbf{LR}} &
  \multicolumn{1}{c|}{\textbf{DT}} &
  \multicolumn{1}{c|}{\textbf{RF}} &
  \multicolumn{1}{c}{\textbf{SVM}} \\ \hline
\multirow{9}{*}{Recall} &
  None &
  \multicolumn{1}{r|}{0} &
  \multicolumn{1}{r|}{0} &
  \multicolumn{1}{r|}{0} &
  \multicolumn{1}{r|}{0} &
  0 &
  \multicolumn{1}{r|}{0} &
  \multicolumn{1}{r|}{0} &
  \multicolumn{1}{r|}{14} &
  \multicolumn{1}{r|}{0} &
  0 &
  \multicolumn{1}{r|}{63} &
  \multicolumn{1}{r|}{0} &
  \multicolumn{1}{r|}{68} &
  \multicolumn{1}{r|}{68} &
  11 \\ 
 &
  RandomOversampler &
  \multicolumn{1}{r|}{0} &
  \multicolumn{1}{r|}{\cellcolor[HTML]{DAE8FC}100} &
  \multicolumn{1}{r|}{0} &
  \multicolumn{1}{r|}{0} &
  \cellcolor[HTML]{DAE8FC}100 &
  \multicolumn{1}{r|}{0} &
  \multicolumn{1}{r|}{57} &
  \multicolumn{1}{r|}{57} &
  \multicolumn{1}{r|}{0} &
  42 &
  \multicolumn{1}{r|}{72} &
  \multicolumn{1}{r|}{65} &
  \multicolumn{1}{r|}{76} &
  \multicolumn{1}{r|}{73} &
  22 \\ 
 &
  SMOTE &
  \multicolumn{1}{r|}{40} &
  \multicolumn{1}{r|}{\cellcolor[HTML]{DAE8FC}100} &
  \multicolumn{1}{r|}{20} &
  \multicolumn{1}{r|}{0} &
  \cellcolor[HTML]{DAE8FC}100 &
  \multicolumn{1}{r|}{0} &
  \multicolumn{1}{r|}{\cellcolor[HTML]{DAE8FC}100} &
  \multicolumn{1}{r|}{42} &
  \multicolumn{1}{r|}{42} &
  0 &
  \multicolumn{1}{r|}{76} &
  \multicolumn{1}{r|}{65} &
  \multicolumn{1}{r|}{78} &
  \multicolumn{1}{r|}{77} &
  22 \\ 
 &
  ADASYN &
  \multicolumn{1}{r|}{60} &
  \multicolumn{1}{r|}{\cellcolor[HTML]{DAE8FC}100} &
  \multicolumn{1}{r|}{0} &
  \multicolumn{1}{r|}{0} &
  \cellcolor[HTML]{DAE8FC}100 &
  \multicolumn{1}{r|}{0} &
  \multicolumn{1}{r|}{\cellcolor[HTML]{DAE8FC}100} &
  \multicolumn{1}{r|}{57} &
  \multicolumn{1}{r|}{42} &
  0 &
  \multicolumn{1}{r|}{78} &
  \multicolumn{1}{r|}{49} &
  \multicolumn{1}{r|}{76} &
  \multicolumn{1}{r|}{77} &
  29 \\ 
 &
  BorderlineSMOTE &
  \multicolumn{1}{r|}{40} &
  \multicolumn{1}{r|}{60} &
  \multicolumn{1}{r|}{0} &
  \multicolumn{1}{r|}{0} &
  60 &
  \multicolumn{1}{r|}{0} &
  \multicolumn{1}{r|}{\cellcolor[HTML]{DAE8FC}100} &
  \multicolumn{1}{r|}{0} &
  \multicolumn{1}{r|}{14} &
  0 &
  \multicolumn{1}{r|}{75} &
  \multicolumn{1}{r|}{49} &
  \multicolumn{1}{r|}{76} &
  \multicolumn{1}{r|}{76} &
  30 \\ 
 &
  SVMSMOTE &
  \multicolumn{1}{r|}{40} &
  \multicolumn{1}{r|}{60} &
  \multicolumn{1}{r|}{0} &
  \multicolumn{1}{r|}{0} &
  40 &
  \multicolumn{1}{r|}{0} &
  \multicolumn{1}{r|}{0} &
  \multicolumn{1}{r|}{0} &
  \multicolumn{1}{r|}{28} &
  0 &
  \multicolumn{1}{r|}{74} &
  \multicolumn{1}{r|}{58} &
  \multicolumn{1}{r|}{79} &
  \multicolumn{1}{r|}{76} &
  23 \\ 
 &
  SMOTUNED &
  \multicolumn{1}{r|}{60} &
  \multicolumn{1}{r|}{\cellcolor[HTML]{DAE8FC}100} &
  \multicolumn{1}{r|}{60} &
  \multicolumn{1}{r|}{60} &
  \cellcolor[HTML]{DAE8FC}100 &
  \multicolumn{1}{r|}{0} &
  \multicolumn{1}{r|}{57} &
  \multicolumn{1}{r|}{57} &
  \multicolumn{1}{r|}{28} &
  57 &
  \multicolumn{1}{r|}{81} &
  \multicolumn{1}{r|}{\cellcolor[HTML]{DAE8FC}100} &
  \multicolumn{1}{r|}{80} &
  \multicolumn{1}{r|}{\cellcolor[HTML]{DAE8FC}83} &
  \cellcolor[HTML]{DAE8FC}100 \\ 
 &
  cWGAN-GP &
  \multicolumn{1}{r|}{80} &
  \multicolumn{1}{r|}{60} &
  \multicolumn{1}{r|}{60} &
  \multicolumn{1}{r|}{80} &
  80 &
  \multicolumn{1}{r|}{28} &
  \multicolumn{1}{r|}{57} &
  \multicolumn{1}{r|}{57} &
  \multicolumn{1}{r|}{43} &
  57 &
  \multicolumn{1}{r|}{79} &
  \multicolumn{1}{r|}{79} &
  \multicolumn{1}{r|}{\cellcolor[HTML]{DAE8FC}83} &
  \multicolumn{1}{r|}{78} &
  49 \\ 
 &
  Dazzle &
  \multicolumn{1}{r|}{\cellcolor[HTML]{DAE8FC}100} &
  \multicolumn{1}{r|}{80} &
  \multicolumn{1}{r|}{\cellcolor[HTML]{DAE8FC}100} &
  \multicolumn{1}{r|}{\cellcolor[HTML]{DAE8FC}100} &
  80 &
  \multicolumn{1}{r|}{\cellcolor[HTML]{DAE8FC}85} &
  \multicolumn{1}{r|}{71} &
  \multicolumn{1}{r|}{\cellcolor[HTML]{DAE8FC}71} &
  \multicolumn{1}{r|}{\cellcolor[HTML]{DAE8FC}57} &
  \cellcolor[HTML]{DAE8FC}71 &
  \multicolumn{1}{r|}{\cellcolor[HTML]{DAE8FC}86} &
  \multicolumn{1}{r|}{84} &
  \multicolumn{1}{r|}{\cellcolor[HTML]{DAE8FC}83} &
  \multicolumn{1}{r|}{\cellcolor[HTML]{DAE8FC}83} &
  78 \\ \hline \hline
\multirow{9}{*}{\begin{tabular}[c]{@{}c@{}}False\\ Positive\\ Rate\end{tabular}} &
  None &
  \multicolumn{1}{r|}{\cellcolor[HTML]{DAE8FC}0} &
  \multicolumn{1}{r|}{\cellcolor[HTML]{DAE8FC}0} &
  \multicolumn{1}{r|}{\cellcolor[HTML]{DAE8FC}0} &
  \multicolumn{1}{r|}{\cellcolor[HTML]{DAE8FC}0} &
  \cellcolor[HTML]{DAE8FC}0 &
  \multicolumn{1}{r|}{\cellcolor[HTML]{DAE8FC}0} &
  \multicolumn{1}{r|}{\cellcolor[HTML]{DAE8FC}0} &
  \multicolumn{1}{r|}{2} &
  \multicolumn{1}{r|}{\cellcolor[HTML]{DAE8FC}0} &
  \cellcolor[HTML]{DAE8FC}0 &
  \multicolumn{1}{r|}{\cellcolor[HTML]{DAE8FC}2} &
  \multicolumn{1}{r|}{\cellcolor[HTML]{DAE8FC}0} &
  \multicolumn{1}{r|}{\cellcolor[HTML]{DAE8FC}4} &
  \multicolumn{1}{r|}{\cellcolor[HTML]{DAE8FC}1} &
  \cellcolor[HTML]{DAE8FC}1 \\ 
 &
  RandomOversampler &
  \multicolumn{1}{r|}{2} &
  \multicolumn{1}{r|}{39} &
  \multicolumn{1}{r|}{1} &
  \multicolumn{1}{r|}{\cellcolor[HTML]{DAE8FC}0} &
  22 &
  \multicolumn{1}{r|}{\cellcolor[HTML]{DAE8FC}0} &
  \multicolumn{1}{r|}{3} &
  \multicolumn{1}{r|}{3} &
  \multicolumn{1}{r|}{\cellcolor[HTML]{DAE8FC}0} &
  7 &
  \multicolumn{1}{r|}{6} &
  \multicolumn{1}{r|}{43} &
  \multicolumn{1}{r|}{6} &
  \multicolumn{1}{r|}{3} &
  2 \\ 
 &
  SMOTE &
  \multicolumn{1}{r|}{17} &
  \multicolumn{1}{r|}{40} &
  \multicolumn{1}{r|}{3} &
  \multicolumn{1}{r|}{\cellcolor[HTML]{DAE8FC}0} &
  24 &
  \multicolumn{1}{r|}{\cellcolor[HTML]{DAE8FC}0} &
  \multicolumn{1}{r|}{96} &
  \multicolumn{1}{r|}{4} &
  \multicolumn{1}{r|}{1} &
  \cellcolor[HTML]{DAE8FC}0 &
  \multicolumn{1}{r|}{9} &
  \multicolumn{1}{r|}{42} &
  \multicolumn{1}{r|}{6} &
  \multicolumn{1}{r|}{4} &
  3 \\ 
 &
  ADASYN &
  \multicolumn{1}{r|}{17} &
  \multicolumn{1}{r|}{41} &
  \multicolumn{1}{r|}{\cellcolor[HTML]{DAE8FC}0} &
  \multicolumn{1}{r|}{\cellcolor[HTML]{DAE8FC}0} &
  25 &
  \multicolumn{1}{r|}{\cellcolor[HTML]{DAE8FC}0} &
  \multicolumn{1}{r|}{96} &
  \multicolumn{1}{r|}{4} &
  \multicolumn{1}{r|}{2} &
  \cellcolor[HTML]{DAE8FC}0 &
  \multicolumn{1}{r|}{12} &
  \multicolumn{1}{r|}{33} &
  \multicolumn{1}{r|}{7} &
  \multicolumn{1}{r|}{6} &
  9 \\ 
 &
  BorderlineSMOTE &
  \multicolumn{1}{r|}{6} &
  \multicolumn{1}{r|}{19} &
  \multicolumn{1}{r|}{\cellcolor[HTML]{DAE8FC}0} &
  \multicolumn{1}{r|}{\cellcolor[HTML]{DAE8FC}0} &
  8 &
  \multicolumn{1}{r|}{\cellcolor[HTML]{DAE8FC}0} &
  \multicolumn{1}{r|}{98} &
  \multicolumn{1}{r|}{2} &
  \multicolumn{1}{r|}{\cellcolor[HTML]{DAE8FC}0} &
  \cellcolor[HTML]{DAE8FC}0 &
  \multicolumn{1}{r|}{9} &
  \multicolumn{1}{r|}{35} &
  \multicolumn{1}{r|}{7} &
  \multicolumn{1}{r|}{5} &
  12 \\ 
 &
  SVMSMOTE &
  \multicolumn{1}{r|}{5} &
  \multicolumn{1}{r|}{11} &
  \multicolumn{1}{r|}{\cellcolor[HTML]{DAE8FC}0} &
  \multicolumn{1}{r|}{\cellcolor[HTML]{DAE8FC}0} &
  7 &
  \multicolumn{1}{r|}{\cellcolor[HTML]{DAE8FC}0} &
  \multicolumn{1}{r|}{\cellcolor[HTML]{DAE8FC}0} &
  \multicolumn{1}{r|}{\cellcolor[HTML]{DAE8FC}1} &
  \multicolumn{1}{r|}{1} &
  \cellcolor[HTML]{DAE8FC}0 &
  \multicolumn{1}{r|}{8} &
  \multicolumn{1}{r|}{42} &
  \multicolumn{1}{r|}{6} &
  \multicolumn{1}{r|}{4} &
  3 \\ 
 &
  SMOTUNED &
  \multicolumn{1}{r|}{19} &
  \multicolumn{1}{r|}{61} &
  \multicolumn{1}{r|}{14} &
  \multicolumn{1}{r|}{20} &
  17 &
  \multicolumn{1}{r|}{\cellcolor[HTML]{DAE8FC}0} &
  \multicolumn{1}{r|}{31} &
  \multicolumn{1}{r|}{3} &
  \multicolumn{1}{r|}{1} &
  14 &
  \multicolumn{1}{r|}{44} &
  \multicolumn{1}{r|}{100} &
  \multicolumn{1}{r|}{35} &
  \multicolumn{1}{r|}{38} &
  98 \\ 
 &
  cWGAN-GP &
  \multicolumn{1}{r|}{17} &
  \multicolumn{1}{r|}{12} &
  \multicolumn{1}{r|}{24} &
  \multicolumn{1}{r|}{20} &
  21 &
  \multicolumn{1}{r|}{2} &
  \multicolumn{1}{r|}{1} &
  \multicolumn{1}{r|}{3} &
  \multicolumn{1}{r|}{1} &
  2 &
  \multicolumn{1}{r|}{17} &
  \multicolumn{1}{r|}{6} &
  \multicolumn{1}{r|}{5} &
  \multicolumn{1}{r|}{17} &
  36 \\ 
 &
  Dazzle &
  \multicolumn{1}{r|}{16} &
  \multicolumn{1}{r|}{20} &
  \multicolumn{1}{r|}{22} &
  \multicolumn{1}{r|}{19} &
  24 &
  \multicolumn{1}{r|}{2} &
  \multicolumn{1}{r|}{2} &
  \multicolumn{1}{r|}{2} &
  \multicolumn{1}{r|}{2} &
  1 &
  \multicolumn{1}{r|}{7} &
  \multicolumn{1}{r|}{11} &
  \multicolumn{1}{r|}{5} &
  \multicolumn{1}{r|}{7} &
  12 \\ \hline \hline
\multirow{9}{*}{G-Measure} &
  None &
  \multicolumn{1}{r|}{0} &
  \multicolumn{1}{r|}{0} &
  \multicolumn{1}{r|}{0} &
  \multicolumn{1}{r|}{0} &
  0 &
  \multicolumn{1}{r|}{0} &
  \multicolumn{1}{r|}{0} &
  \multicolumn{1}{r|}{24} &
  \multicolumn{1}{r|}{0} &
  0 &
  \multicolumn{1}{r|}{77} &
  \multicolumn{1}{r|}{0} &
  \multicolumn{1}{r|}{80} &
  \multicolumn{1}{r|}{80} &
  19 \\ 
 &
  RandomOversampler &
  \multicolumn{1}{r|}{0} &
  \multicolumn{1}{r|}{75} &
  \multicolumn{1}{r|}{0} &
  \multicolumn{1}{r|}{0} &
  87 &
  \multicolumn{1}{r|}{0} &
  \multicolumn{1}{r|}{71} &
  \multicolumn{1}{r|}{71} &
  \multicolumn{1}{r|}{0} &
  58 &
  \multicolumn{1}{r|}{81} &
  \multicolumn{1}{r|}{61} &
  \multicolumn{1}{r|}{84} &
  \multicolumn{1}{r|}{83} &
  36 \\ 
 &
  SMOTE &
  \multicolumn{1}{r|}{53} &
  \multicolumn{1}{r|}{74} &
  \multicolumn{1}{r|}{33} &
  \multicolumn{1}{r|}{0} &
  86 &
  \multicolumn{1}{r|}{0} &
  \multicolumn{1}{r|}{6} &
  \multicolumn{1}{r|}{59} &
  \multicolumn{1}{r|}{59} &
  0 &
  \multicolumn{1}{r|}{83} &
  \multicolumn{1}{r|}{61} &
  \multicolumn{1}{r|}{85} &
  \multicolumn{1}{r|}{85} &
  36 \\ 
 &
  ADASYN &
  \multicolumn{1}{r|}{69} &
  \multicolumn{1}{r|}{74} &
  \multicolumn{1}{r|}{0} &
  \multicolumn{1}{r|}{0} &
  85 &
  \multicolumn{1}{r|}{0} &
  \multicolumn{1}{r|}{7} &
  \multicolumn{1}{r|}{71} &
  \multicolumn{1}{r|}{59} &
  0 &
  \multicolumn{1}{r|}{82} &
  \multicolumn{1}{r|}{56} &
  \multicolumn{1}{r|}{84} &
  \multicolumn{1}{r|}{85} &
  43 \\ 
 &
  BorderlineSMOTE &
  \multicolumn{1}{r|}{56} &
  \multicolumn{1}{r|}{68} &
  \multicolumn{1}{r|}{0} &
  \multicolumn{1}{r|}{0} &
  72 &
  \multicolumn{1}{r|}{0} &
  \multicolumn{1}{r|}{3} &
  \multicolumn{1}{r|}{0} &
  \multicolumn{1}{r|}{24} &
  0 &
  \multicolumn{1}{r|}{82} &
  \multicolumn{1}{r|}{55} &
  \multicolumn{1}{r|}{83} &
  \multicolumn{1}{r|}{84} &
  45 \\ 
 &
  SVMSMOTE &
  \multicolumn{1}{r|}{56} &
  \multicolumn{1}{r|}{71} &
  \multicolumn{1}{r|}{0} &
  \multicolumn{1}{r|}{0} &
  55 &
  \multicolumn{1}{r|}{0} &
  \multicolumn{1}{r|}{0} &
  \multicolumn{1}{r|}{0} &
  \multicolumn{1}{r|}{44} &
  0 &
  \multicolumn{1}{r|}{82} &
  \multicolumn{1}{r|}{57} &
  \multicolumn{1}{r|}{85} &
  \multicolumn{1}{r|}{84} &
  37 \\ 
 &
  SMOTUNED &
  \multicolumn{1}{r|}{68} &
  \multicolumn{1}{r|}{55} &
  \multicolumn{1}{r|}{70} &
  \multicolumn{1}{r|}{68} &
  \cellcolor[HTML]{DAE8FC}90 &
  \multicolumn{1}{r|}{0} &
  \multicolumn{1}{r|}{62} &
  \multicolumn{1}{r|}{71} &
  \multicolumn{1}{r|}{44} &
  68 &
  \multicolumn{1}{r|}{69} &
  \multicolumn{1}{r|}{0} &
  \multicolumn{1}{r|}{75} &
  \multicolumn{1}{r|}{73} &
  3 \\ 
 &
  cWGAN-GP &
  \multicolumn{1}{r|}{81} &
  \multicolumn{1}{r|}{71} &
  \multicolumn{1}{r|}{67} &
  \multicolumn{1}{r|}{80} &
  79 &
  \multicolumn{1}{r|}{44} &
  \multicolumn{1}{r|}{72} &
  \multicolumn{1}{r|}{72} &
  \multicolumn{1}{r|}{59} &
  72 &
  \multicolumn{1}{r|}{81} &
  \multicolumn{1}{r|}{\cellcolor[HTML]{DAE8FC}86} &
  \multicolumn{1}{r|}{\cellcolor[HTML]{DAE8FC}88} &
  \multicolumn{1}{r|}{80} &
  56 \\ 
 &
  Dazzle &
  \multicolumn{1}{r|}{\cellcolor[HTML]{DAE8FC}91} &
  \multicolumn{1}{r|}{\cellcolor[HTML]{DAE8FC}79} &
  \multicolumn{1}{r|}{\cellcolor[HTML]{DAE8FC}87} &
  \multicolumn{1}{r|}{\cellcolor[HTML]{DAE8FC}89} &
  79 &
  \multicolumn{1}{r|}{\cellcolor[HTML]{DAE8FC}91} &
  \multicolumn{1}{r|}{\cellcolor[HTML]{DAE8FC}83} &
  \multicolumn{1}{r|}{\cellcolor[HTML]{DAE8FC}82} &
  \multicolumn{1}{r|}{\cellcolor[HTML]{DAE8FC}72} &
  \cellcolor[HTML]{DAE8FC}83 &
  \multicolumn{1}{r|}{\cellcolor[HTML]{DAE8FC}89} &
  \multicolumn{1}{r|}{\cellcolor[HTML]{DAE8FC}86} &
  \multicolumn{1}{r|}{\cellcolor[HTML]{DAE8FC}88} &
  \multicolumn{1}{r|}{\cellcolor[HTML]{DAE8FC}87} &
  \cellcolor[HTML]{DAE8FC}83 \\ \hline \hline
\multirow{9}{*}{F-Measure} &
  None &
  \multicolumn{1}{r|}{0} &
  \multicolumn{1}{r|}{0} &
  \multicolumn{1}{r|}{0} &
  \multicolumn{1}{r|}{0} &
  0 &
  \multicolumn{1}{r|}{0} &
  \multicolumn{1}{r|}{0} &
  \multicolumn{1}{r|}{9} &
  \multicolumn{1}{r|}{0} &
  0 &
  \multicolumn{1}{r|}{71} &
  \multicolumn{1}{r|}{0} &
  \multicolumn{1}{r|}{68} &
  \multicolumn{1}{r|}{\cellcolor[HTML]{DAE8FC}77} &
  18 \\ 
 &
  RandomOversampler &
  \multicolumn{1}{r|}{0} &
  \multicolumn{1}{r|}{4} &
  \multicolumn{1}{r|}{0} &
  \multicolumn{1}{r|}{0} &
  7 &
  \multicolumn{1}{r|}{0} &
  \multicolumn{1}{r|}{26} &
  \multicolumn{1}{r|}{29} &
  \multicolumn{1}{r|}{0} &
  12 &
  \multicolumn{1}{r|}{67} &
  \multicolumn{1}{r|}{27} &
  \multicolumn{1}{r|}{69} &
  \multicolumn{1}{r|}{74} &
  31 \\ 
 &
  SMOTE &
  \multicolumn{1}{r|}{3} &
  \multicolumn{1}{r|}{4} &
  \multicolumn{1}{r|}{\cellcolor[HTML]{DAE8FC}8} &
  \multicolumn{1}{r|}{0} &
  6 &
  \multicolumn{1}{r|}{0} &
  \multicolumn{1}{r|}{2} &
  \multicolumn{1}{r|}{18} &
  \multicolumn{1}{r|}{35} &
  0 &
  \multicolumn{1}{r|}{62} &
  \multicolumn{1}{r|}{27} &
  \multicolumn{1}{r|}{69} &
  \multicolumn{1}{r|}{74} &
  31 \\ 
 &
  ADASYN &
  \multicolumn{1}{r|}{5} &
  \multicolumn{1}{r|}{4} &
  \multicolumn{1}{r|}{0} &
  \multicolumn{1}{r|}{0} &
  6 &
  \multicolumn{1}{r|}{0} &
  \multicolumn{1}{r|}{2} &
  \multicolumn{1}{r|}{23} &
  \multicolumn{1}{r|}{26} &
  0 &
  \multicolumn{1}{r|}{58} &
  \multicolumn{1}{r|}{25} &
  \multicolumn{1}{r|}{67} &
  \multicolumn{1}{r|}{70} &
  29 \\ 
 &
  BorderlineSMOTE &
  \multicolumn{1}{r|}{8} &
  \multicolumn{1}{r|}{4} &
  \multicolumn{1}{r|}{0} &
  \multicolumn{1}{r|}{0} &
  \cellcolor[HTML]{DAE8FC}10 &
  \multicolumn{1}{r|}{0} &
  \multicolumn{1}{r|}{2} &
  \multicolumn{1}{r|}{0} &
  \multicolumn{1}{r|}{22} &
  0 &
  \multicolumn{1}{r|}{61} &
  \multicolumn{1}{r|}{24} &
  \multicolumn{1}{r|}{67} &
  \multicolumn{1}{r|}{71} &
  27 \\ 
 &
  SVMSMOTE &
  \multicolumn{1}{r|}{9} &
  \multicolumn{1}{r|}{8} &
  \multicolumn{1}{r|}{0} &
  \multicolumn{1}{r|}{0} &
  7 &
  \multicolumn{1}{r|}{0} &
  \multicolumn{1}{r|}{0} &
  \multicolumn{1}{r|}{0} &
  \multicolumn{1}{r|}{26} &
  0 &
  \multicolumn{1}{r|}{64} &
  \multicolumn{1}{r|}{25} &
  \multicolumn{1}{r|}{70} &
  \multicolumn{1}{r|}{73} &
  32 \\ 
 &
  SMOTUNED &
  \multicolumn{1}{r|}{5} &
  \multicolumn{1}{r|}{2} &
  \multicolumn{1}{r|}{6} &
  \multicolumn{1}{r|}{4} &
  8 &
  \multicolumn{1}{r|}{0} &
  \multicolumn{1}{r|}{4} &
  \multicolumn{1}{r|}{30} &
  \multicolumn{1}{r|}{22} &
  \cellcolor[HTML]{DAE8FC}85 &
  \multicolumn{1}{r|}{36} &
  \multicolumn{1}{r|}{21} &
  \multicolumn{1}{r|}{40} &
  \multicolumn{1}{r|}{39} &
  22 \\ 
 &
  cWGAN-GP &
  \multicolumn{1}{r|}{7} &
  \multicolumn{1}{r|}{\cellcolor[HTML]{DAE8FC}9} &
  \multicolumn{1}{r|}{4} &
  \multicolumn{1}{r|}{6} &
  6 &
  \multicolumn{1}{r|}{19} &
  \multicolumn{1}{r|}{\cellcolor[HTML]{DAE8FC}50} &
  \multicolumn{1}{r|}{32} &
  \multicolumn{1}{r|}{35} &
  40 &
  \multicolumn{1}{r|}{53} &
  \multicolumn{1}{r|}{\cellcolor[HTML]{DAE8FC}70} &
  \multicolumn{1}{r|}{\cellcolor[HTML]{DAE8FC}76} &
  \multicolumn{1}{r|}{51} &
  24 \\ 
 &
  Dazzle &
  \multicolumn{1}{r|}{\cellcolor[HTML]{DAE8FC}10} &
  \multicolumn{1}{r|}{6} &
  \multicolumn{1}{r|}{7} &
  \multicolumn{1}{r|}{\cellcolor[HTML]{DAE8FC}8} &
  5 &
  \multicolumn{1}{r|}{\cellcolor[HTML]{DAE8FC}52} &
  \multicolumn{1}{r|}{\cellcolor[HTML]{DAE8FC}50} &
  \multicolumn{1}{r|}{\cellcolor[HTML]{DAE8FC}42} &
  \multicolumn{1}{r|}{\cellcolor[HTML]{DAE8FC}42} &
  55 &
  \multicolumn{1}{r|}{\cellcolor[HTML]{DAE8FC}72} &
  \multicolumn{1}{r|}{64} &
  \multicolumn{1}{r|}{\cellcolor[HTML]{DAE8FC}76} &
  \multicolumn{1}{r|}{70} &
  \cellcolor[HTML]{DAE8FC}58 \\ \hline \hline
\end{tabular}
\label{tbl:results}
\end{table*}

Our study answers the following research questions:

\begin{RQ}
{\bf RQ1.} Will GANs based oversampling better than SMOTE based oversampling?
\end{RQ}

For each treatment in our study, we use default learners for fair comparison. Table~\ref{tbl:results} lists all the evaluations results of metrics defined in Table~\ref{tbl:metric} for all three datasets. In these results, the \textit{None} treatment indicate the training process with original dataset without any oversampling techniques, after which different oversamplers such as \textit{RandomOversampler} and variants of SMOTE are presented. The \textit{cWGAN-GP} treatment is GANs based oversampler with optimization. In order to configure the architecture of cWGAN-GP, we randomly select parameter set from Table~\ref{tbl:DazzleParameter} during each run. 

As we can observe from the table, the original dataset without oversampling performs badly across all datasets, even with different machine learning algorithms. The results are no surprise as we consider the percentage of security relevant class samples in Table~\ref{tbl:dataset}. For moodle and Ambari dataset, the positive class samples are less than 3\% of the whole datasets, it is hard for machine learning algorithms to learn the traits with so few samples. As a result, none of the learners can detect any true positive during the testing phase. 

Naive oversampler such as the RandomOversampler shows some advantages for some learners, for example Logistic Regression and SVM, but fails for others. SMOTE and its variants demonstrate better results than RandomOversampler, but the advantage is not obvious. Previous state-of-the-art SMOTUNED works best among all SMOTE based oversampling techniques.

cWGAN-GP is the GANs version oversampler, and there are two observations from the results:
\begin{itemize}
    \item cWGAN-GP achieves nearly tied performance with SMOTUNED in important metric such as recall, but we note that the latter is an optimized version which requires more effort and configuration.
    \item Unlike other oversamplers, cWGAN-GP does not fail totally in some certain machine learning algorithms. For example, the SVMSMOTE does not detect any true positive with LR and RF in Moodle dataset, and even 4 out of 5 learners fails in Ambari dataset. This phenomenon indicates that cWGAN-GP is more practical to use in general cases. 
\end{itemize}

We have to point out that the false positive rate metric is not suggested to indicate which method is better than others. For example, in Table~\ref{tbl:results}, for the Random Forest results of Moodle vulnerable files dataset, several treatments have achieve zero false positive rate (therefore highlighted in blue color), however, their recall results are also zero. Thus, these treatments are not recommended.

Since Dazzle optimizes GANs with the goal of increasing g-measure, which is the harmonic mean of recall and the complement of false positive rate. Therefore, when the g-measure is increased, there would be three cases: 1) recall increased, FPR decreased; 2) recall increased, FPR increased; and 3) recall decreased, FPR decreased. Our results would fall into these three groups. When we notice that some improvements in recall come at the cost of increments in false positive rate, while the ideal false positive rate is zero. We say that the trade-off between the increments of recall and false positive rate is still acceptable, especially in mission-critical security tasks, as we do not want to miss any security relevant target samples in the detection. Such a ``price'' indicates more extra effort to read more source code or bug reports for security practitioners, and it is the price of software quality assurance.

\begin{bclogo}[logo={\bccrayon},arrondi=0.1]{Answer}
cWGAN-GP is more practical to use in  the general cases, as it is not sensitive to certain machine learning algorithms.
\end{bclogo}
 
\begin{RQ}
{\bf RQ2.} Will Dazzle (optimized GAN) work even better?
\end{RQ}

We optimize Dazzle with the goal of g-measure, which ideally with high recall and lower false positive rate. As we can observe from the result table, Dazzle works even better than cWGAN-GP. Benefit from optimizing, Dazzle achieves an average of improvement rate of 30\%, 62\% and 17\% over cWGAN-GP in recall, respectively. This is explainable, as the ``default'' (with randomly selection in our case) hyperparameters for GANs might bring in issues in Section 3, hence is not one-size-fits-all across all scenarios and should be deprecated. We would recommend exploring and developing specialized tools for certain local domain.

\begin{bclogo}[logo={\bccrayon},arrondi=0.1]{Answer}
Dazzle (the optimized version) shows even better performance than cWGAN-GP across all studied datasets.
\end{bclogo}

\begin{RQ}
{\bf RQ3.} Is optimized GANs (i.e., Dazzle) impractically slow?
\end{RQ}

As shown in \textbf{RQ1}, Dazzle achieves promising improvement over baseline treatments in performance with 30 iterations of optimization trails. Table~\ref{tbl:runtime} lists the average runtime of each treatment of all machine learning algorithms. Considering the complexity of optimizing the architecture of neural networks, Dazzle is not surprisingly takes the most runtime cost. However, considering the mission-critical nature of the security tasks we are addressing, we would comment that the trade-off between performance and runtime is still worth. The experiment is carried out with CPU resources only, and with the help of GPU or parallel computing, Dazzle could be configured to be more practical to use.

\begin{bclogo}[logo={\bccrayon},arrondi=0.1]{Answer}
Even with more runtime, Dazzle still worths the trade-off when considering the improvement in performance.
\end{bclogo}

% Lastly, we make two recommendations. Firstly, at the operational level, we recommend to try Dazzle for other security tasks that suffer from class imbalance issues. Secondly, at the methodological level, we recommend:

% \begin{itemize}
%     \item Analysts should select solutions that fit to their problem space. Existing general methods might not always be the optimal choice for specific tasks. 
%     \item Always try hyperparameter optimization. Methods that seem less than useful in their ``off-the-shelf'' configurations can become later state-of-the-art after tuning.
% \end{itemize}

Lastly, we believe that there are several directions that can be explored after this work. For example, we would like to try more security tasks to check and endorse the merits of the proposed methods in other cases. Secondly, we would plan to compare with other baselines such as recent improvements on SMOTE and methods other than oversampling. Thirdly, we would like to perform more analysis of the new samples generated by the proposed method to get a better understanding of the methods.

\section{Threats to Validity}\label{sec:threats}

% In this section, we discuss the possible factors that affect the effectiveness of evaluations. Such factors also commonly exist in other research works with large scale empirical studies.

\subsection{Evaluation Bias} In our work, we choose some commonly used metrics for evaluation purpose and set \textit{g-measure} as our optimization target. We do not use some other metrics because relevant information is not available to us or we think they are not suitable enough to this specific task (e.g., precision). In addition, we use equal weight in recall and specificity in the definition of g-measure, which is widely adopted in existing literature. We agree that it is important for these two elements to be re-weighted for different tasks, and this can be further explored as one of our future directions. Our implementation is flexible and we can adjust to proper metrics or balances with minor code modification.

\subsection{Parameter Bias} We have to note that default hyperparameter values have been used for the baseline machine learning algorithms, which means that the performance results reported in Table~\ref{tbl:results} might be suboptimal for baseline methods. To some degree, this also might have the effect of magnifying the advantages of the proposed method. Previous studies have also indicated that it is a good practice to avoid using default settings of machine learning algorithms~\cite{kudjo2019significant}~\cite{shu2021better}~\cite{tantithamthavorn2018impact}. In the case if those hyperparameters have been tuned, the conclusions from the proposed method might be different.

\subsection{Learner Bias} Research into automatic classifiers is a large and active field. While different machine learning algorithms have been developed to solve different classification problem tasks. Any data mining study, such as this paper, can only use a small subset of the known classification algorithms. For this work, we select machine learning algorithms that are commonly used in classification tasks.

\subsection{Input Bias} Our results come from the space of hyperparameter optimization explored in this paper. In theory, other ranges might lead to other results. That said, our goal here is not to offer the {\em best} optimization but to argue that optimized GANs architecture is better than current state-of-the-art oversampler in addressing class imbalance. For those purposes, we would argue that our current results suffice.

\subsection{Dataset Bias} This empirical study demonstrates the effectiveness of the proposed method in security vulnerability/bug report datasets. However, the internal difference between studied datasets and datasets from other security tasks (e.g., in Table~\ref{tbl:dataAugumentation}) or from other domains cannot be ignored. Therefore, there is no guarantee that the findings in this study would still hold in other datasets.

\begin{table}[!t]
\small
\centering
\caption{Runtime of oversampling treatments in minutes for each dataset. Note that ``<'' means the runtime is close but less than the given results.}
\begin{tabular}{l|c|c|c}
\hline
\multicolumn{1}{c|}{Treatment} &
  \begin{tabular}[c]{@{}c@{}}Moodle\\ Vulnerable Files\end{tabular} &
  \begin{tabular}[c]{@{}c@{}}Ambari\\ Bug Report\end{tabular} &
  \begin{tabular}[c]{@{}c@{}}JavaScript\\ Vulnerability\end{tabular} \\ \hline
RandomOversampler & < 1 & < 1 & < 1 \\ 
SMOTE             & < 1 & < 1 & < 1 \\ 
ADASYN            & < 1 & < 1 & < 1 \\ 
BorderlineSMOTE   & < 1 & < 1 & < 1 \\ 
SVMSMOTE          & < 1 & < 1 & < 1 \\ 
SMOTUNED          & < 3 & < 2 & < 5 \\ 
cWGAN-GP          & < 5 & < 5 & < 5 \\ 
Dazzle            & < 25 & < 25 & < 30 \\ \hline
\end{tabular}
\label{tbl:runtime}
\end{table} 
\section{Other Notes on Class Imbalance}~\label{sec:discussion}
Class-imbalance learning~\cite{liu2008exploratory} refers to methods to hancle class imbalance issues. Data oversampling is not the only effective way to address data imbalancing issues. Other approaches can mainly fall into the following categorizations according the the problem space:

\textbf{\textit{Data Sampling level.}} Apart from data oversampling, data undersampling~\cite{lin2017clustering,koziarski2020radial} is another alternative way to deal with class imbalance from the data level. In data undersampling, we can remove some instances from majority class. Generally, this method is suggested when there is large number of training instances. However, data undersampling might suffer from information loss due to removal of majority class instances.

\textbf{\textit{Model Training level.}} Many prior work propose various ways to train efficient models with class imbalanced datasets. For example, bagging ensemble~\cite{galar2011review} is a technique that divides the original training datasets into several subsets of same size, while each subset is used to train a single classifier, and then the method aggregates individual classifiers into an ensemble classifier. This method is well-known for its simplicity and good generalization ability. Some other work~\cite{shu2021better} applies hyperparameter optimization on both data pre-processors and machine learning models to explore optimal hyperparameter settings that work for imbalanced data. Several studies use cost-sensitive learning~\cite{cao2013pso}~\cite{cao2013optimized} in the field of imbalanced learning. Cost-sensitive learning takes the costs prediction errors into consideration and does not treat all classification errors as equal. This makes sense in some security  scenarios, for example, classifying a benign software as a malware (i.e., false positive case) is less of a problem than classifying a malware case as a benign software (i.e., false negative case) since security practitioners do not hope to miss any malware sample. Furthermore, transfer learning~\cite{al2016transfer} is another promising technique that use auxiliary data to augment learning when training samples are not sufficient. This algorithm works by including a similar and possibly larger dataset, with which perform knowledge transfer.

\textbf{\textit{Feature Selection level.}} Using feature-selection for addressing class imbalance is not a largely explored research area compared to previous levels.  Several few work investigate feature selection with imbalanced data empirically~\cite{grobelnik1999feature,zheng2004feature}, and researchers~\cite{ali2013classification} warns that the extra computational cost would be an issue of concern.

We note that the \textit{\textbf{scope of this work}} is not to explore and compare all data imbalance solutions extensively. It is not fair to offer a general conclusion that one technique outperforms other techniques in all tasks. Rather, the focus of this work is to explore the merits of optimized generative adversarial network as an data oversampling technique in addressing security dataset imbalance issues. A hybrid combination of above mentioned approaches (including this work) might work even better, and we would like to explore as an interesting future direction.

\section{Conclusion}\label{sec:conclusion}

When the target class is rare, as it is often within security datasets, it is hard for a machine learning algorithm to distinguish the goal (security target) from others (the normal events). To address such class imbalance issue, prior researchers in software engineering often use SMOTE (or its variants, see Table~\ref{tbl:smoteVairant}) as a solution. SMOTE was first proposed in 2002, nearly two decades ago. This paper seeks a better method than SMOTE. 

One recent alternative to SMOTE is the Generative  Adversarial Networks. This architecture contains two components (the generator and the discriminator)  that ``fight it out'' to generate new examples. The experience has been that it is hard to balance these two components manually, so we experimented with addressing that problem with automatic hyperparameter optimization.

The empirical study shows that GANs with hyperparameter optimization outperforms prior SMOTE (and its variants) and standard GANs (without optimization). For example, Dazzle can achieve an average of about 60\% improvement rate over SMOTE in recall on studied dataset among different classifiers. Based on this study, we recommend using GANs with hyperparameter optimization (and not off-the-shelf default settings) to train a good security vulnerability prediction model (from the view of data oversampling). More generally, we suggest using hyperparameter optimization in other tasks in SE community.

\balance
\bibliographystyle{ACM-Reference-Format}
\bibliography{main.bbl}

\end{document}